\documentclass[journal]{vgtc}                     % final (journal style)

\usepackage{soul,color}
\usepackage[ruled,linesnumbered]{algorithm2e}
\usepackage{amsfonts}
\usepackage{amsmath}
\usepackage{amssymb}
\usepackage{amsthm}
\usepackage{multirow,multicol}
\usepackage{pifont}
\onlineid{1290}

\vgtccategory{Research}

\vgtcpapertype{application/design study}

\newcommand{\systemname}{\textit{HealthPrism}\xspace}
\newcommand{\ie}{i.e.}
\newcommand{\eg}{e.g.}
\newcommand{\etal}{et al.}

\soulregister\cite7
\soulregister\ref7
\soulregister\pageref7
\soulregister\figurename7
\soulregister\subsection7
\soulregister\tablename7
\soulregister\systemname7
\soulregister\bfseries7
\soulregister\textbf7

\title{\systemname: A Visual Analytics System for Exploring Children's Physical and Mental Health Profiles with Multimodal Data}

\author{%
  \authororcid{Zhihan Jiang}{0000-0003-4857-7143},
  \authororcid{Handi Chen}{0000-0002-4223-3502}, \authororcid{Rui Zhou}{0009-0002-3229-0324}, \authororcid{Jing Deng}{0009-0002-8935-4157}, \authororcid{Xinchen Zhang}{0000-0003-3650-7332}, \authororcid{Running Zhao}{0000-0003-2496-3429}, Cong Xie, \\\authororcid{Yifang Wang*}{0000-0001-6267-9440}, and \authororcid{Edith C.H. Ngai*}{0000-0002-3454-8731}
}

\authorfooter{
  \item
  	Z. Jiang, H. Chen, R. Zhou, J. Deng, X. Zhang, R. Zhao, and E. Ngai are with the University of Hong Kong.
  	E-mail: \{zhjiang, hdchen, zackery, u3008395, u3008407, rnzhao\}@connect.hku.hk, chngai@eee.hku.hk.
  \item
  	C. Xie is with Tencent.
  	E-mail: xie.cong@outlook.com.

   \item Y. Wang is with Kellogg School of Management, Northwestern University
   E-mail: yifang.wang@kellogg.northwestern.edu.

  \item * Co-corresponding authors.
}

\abstract{The correlation between children's personal and family characteristics (\eg, demographics and socioeconomic status) and their physical and mental health status has been extensively studied across various research domains, such as public health, medicine, and data science. 
Such studies can provide insights into the underlying factors affecting children's health and aid in the development of targeted interventions to improve their health outcomes. 
However, with the availability of multiple data sources, including context data (\ie, the background information of children) and motion data (\ie, sensor data measuring activities of children), new challenges have arisen due to the large-scale, heterogeneous, and multimodal nature of the data. 
Existing statistical hypothesis-based and learning model-based approaches have been inadequate for comprehensively analyzing the complex correlation between multimodal features and multi-dimensional health outcomes due to the limited information revealed. 
In this work, we first distill a set of design requirements from multiple levels through conducting a literature review and iteratively interviewing 11 experts from multiple domains (\eg, public health and medicine). 
Then, we propose \systemname, an interactive visual and analytics system for assisting researchers in exploring the importance and influence of various context and motion features on children's health status from multi-level perspectives. 
Within \systemname, a multimodal learning model with a gate mechanism is proposed for health profiling and cross-modality feature importance comparison. 
A set of visualization components is designed for experts to explore and understand multimodal data freely.
We demonstrate the effectiveness and usability of \systemname through quantitative evaluation of the model performance, case studies, and expert interviews in associated domains.}

\teaser{
  \centering
  \includegraphics[width=0.94\linewidth]{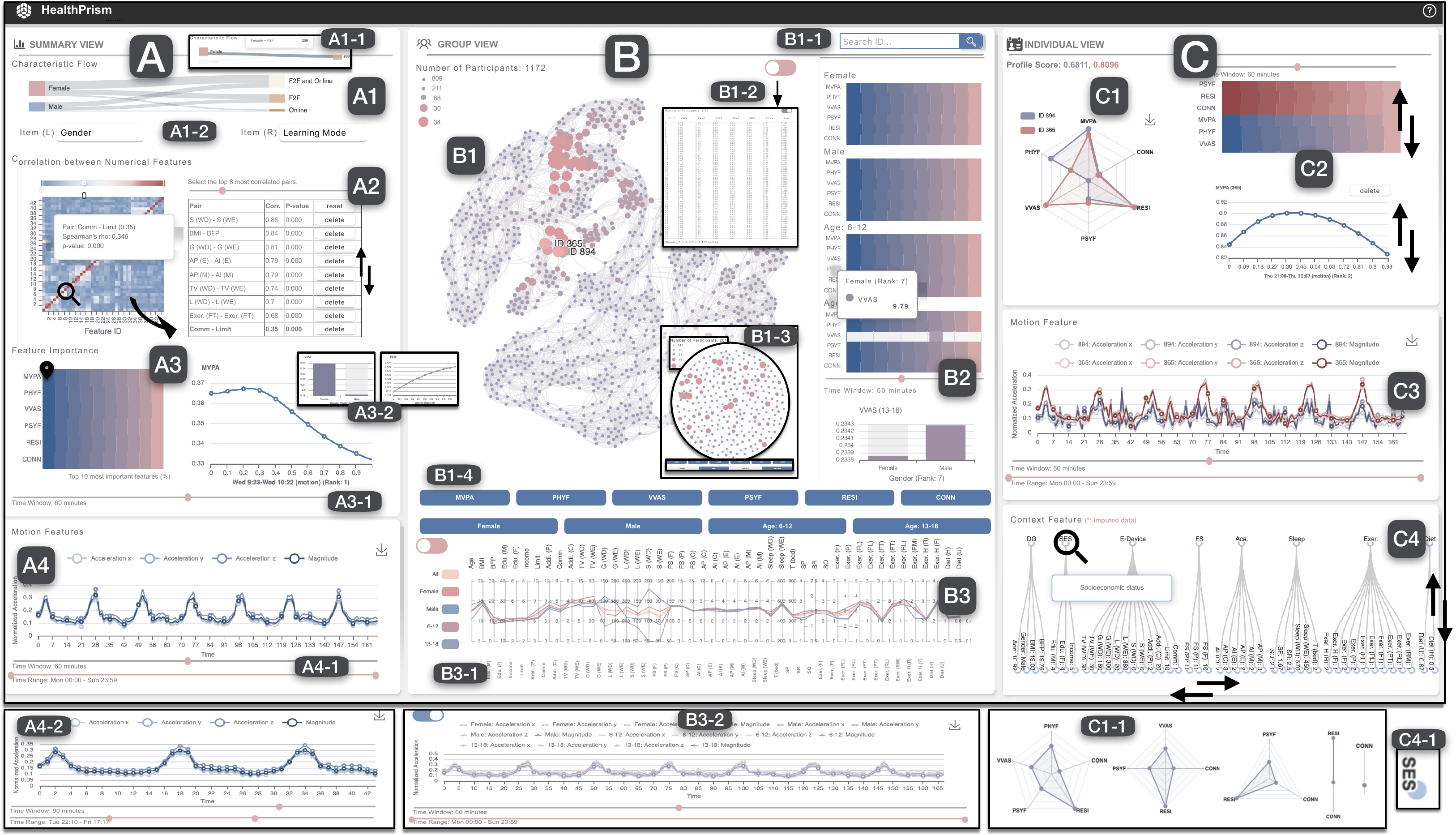}
  \caption{%
  	The overview of \systemname. The Summary View (A) presents the overall context and motion features, including the flows of categorical context features (A1), the correlation between numerical context features (A2), the distribution of motion features (A4), and the importance and influence of context and motion features (A3). The Group View (B) presents a network graph (B1) showing the clusters of health profiles of participants according to the health indicators, genders, and age groups selected by users. It also presents the feature importance and influence (B2) and feature overview (B3) with different groups. By clicking on the nodes in the graph, the Individual View (C) will display the health profile (C1), motion feature (C3), and context feature (C4) of corresponding individuals, as well as the feature importance and influence (C2). The Individual View is able to display a maximum of two individuals for comparison.%
  }
  \label{fig:teaser}
  \vspace{-0.1cm}
}

\usepackage{tabu}                      % only used for the table example
\usepackage{booktabs}                  % only used for the table example
\usepackage{lipsum}                    % used to generate placeholder text
\usepackage{mwe}                       % used to generate placeholder figures
\usepackage{setspace}
\usepackage{mathptmx}                  % use matching math font

\begin{document}
\begin{spacing}{0.97} %{0.97}

%%%%%%%%%%%%%%%%%%%%%%%%%%%%%%%%%%%%%%%%%%%%%%%%%%%%%%%%%%%%%%%%
%%%%%%%%%%%%%%%%%%%%%% START OF THE PAPER %%%%%%%%%%%%%%%%%%%%%%
%%%%%%%%%%%%%%%%%%%%%%%%%%%%%%%%%%%%%%%%%%%%%%%%%%%%%%%%%%%%%%%%

\firstsection{Introduction}

\maketitle

% General Background: Provide some background information on the importance of health profiling, and briefly introduce the specific population that your study focuses on (e.g. children).

Children's physical and mental health are correlated with their personal and family background (\eg, family socioeconomic status \cite{bradley2002socioeconomic}, academic performance \cite{polanin2021meta}, and physical activity patterns \cite{world2008pacific}). Researchers in various domains, such as public health, medicine, and data science, have proposed numerous works studying the patterns and relationships between these background characteristics and the health outcomes to provide insights into the underlying factors that affect children's health status and develop target interventions to improve their health \cite{national2000health}. 

In recent years, the prevalence of wearable devices with multiple sensors has enabled the collection of large-scale continuous sensor signals from children. With the availability of multiple data sources, including context data (i.e., the background characteristics of children) and motion data (i.e., sensor data measuring activities of children), existing \textcolor{black}{statistical hypothesis}-based and \textcolor{black}{learning model}-based approaches have been inadequate due to the large-scale, heterogeneous, and multimodal data involved. \textcolor{black}{The statistical hypothesis-based methods often rely on inflexible assumptions \cite{kaplan2001demographic, polanin2021meta, lee2020associations}, which may not be compatible with multimodal features}. Others are based on \textcolor{black}{learning models} and explore the importance and influence of features on health outcomes by interpreting the models \cite{miotto2018deep, waring2020automated, healthinference}. However, many traditional machine learning models are also unable to deal with multimodal data, and the models based on deep learning usually have limited interpretability \cite{shorten2021deep}. Furthermore, building machine learning or deep learning models requires specialized technical expertise, which may be non-trivial for many researchers in public health and medical domains to perceive correlations between the input and output of the models \cite{dudley2018review}.

Visual analytics is a promising approach for \textcolor{black}{effectively} analyzing multimodal data to profile children's health. However, there are few studies from the visualization community that have explored this area, specifically utilizing context and motion patterns from multi-level views \textcolor{black}{(i.e., overall, group, and individual levels)}. Most previous research in healthcare visualization has focused on analyzing electronic health records \cite{zhang2013five, hakone2016proact}, temporal events for patients \cite{loorak2015timespan,bernard2018using, guo2018visual}, and healthcare dashboards \cite{elshehaly2020qualdash, kwon2015visohc, zhuang2022framework}. Although Cheng \etal \cite{cheng2021vbridge} built a visual analytics tool to support clinicians' decision-making workflow, their analysis is limited to tabular data. Therefore, the aim of this study is to develop an interactive visualization system for exploring children's health profiles with multimodal features. 

To this end, there are several challenges to be addressed. First, it is difficult to visualize a large volume of multimodal data in a comprehensive and understandable way. Second, it is non-trivial to depict the correlation between the different types of input (\ie, context and motion features) and multiple health outcomes. Furthermore, comparing the importance of features with different modalities is also challenging \cite{chen2019synergistic}. Third, supporting multi-level analysis and reasoning of multimodal data is difficult, as presenting patterns from different perspectives and identifying targets for further investigation requires substantial effort.

% Second, it is non-trivial to depict the correlation between multimodal input and multi-dimensional output. 
% Children's context features are tabular data, and motion features are time series data, and there are multiple outcomes. 
% The features have different importance and influence on different health outcomes. 
% The comparison of importance between features with different modalities is also challenging \cite{chen2019synergistic}. 
% Third, supporting multi-level analysis and reasoning of multimodal data is difficult. 
% Presenting the patterns from the perspectives of groups and individuals for intuitive comparison and identifying targets to be further investigated requires substantial effort.

In this work, we first review the literature and collaborate with 11 experts from multiple domains (e.g., public health and medicine) to distill multi-level. \textcolor{black}{Based on the context and motion data collected from 1,172 children in Hong Kong and their corresponding multi-dimensional health profiles (including six indicators, i.e., physical activity intensity, physical functioning, health confidence, psychosocial functioning, resilience, and connectedness), we propose a multimodal learning model with a gate mechanism that can automatically adjust feature weights to infer the health indicators. Then, we propose \systemname~(\figurename{~\ref{fig:teaser}}), an interactive visual analytics system consisting of \emph{Summary View}, \emph{Group View}, and \emph{Individual View} for exploring multimodal feature importance and influence on children's physical and mental health profiles.}

% utilizing contextual features and motion data collected from 1,172 children in Hong Kong, along with their corresponding multi-dimensional health profiles, we propose \systemname, an interactive visual analytics system consisting of three views (Summary View, Group View, Individual View). This system assists researchers in exploring children's health profiles with multimodal features from multiple levels.

To summarize, our major contributions are:

\begin{itemize}
    \item \systemname, an interactive visual analytics system to support researchers in associated domains to explore children's health profiles with multimodal features \textcolor{black}{effectively} and efficiently. 
    % Such an intuitive and effective system can provide insights into the underlying factors that affect children's health status.
    \item A multimodal learning model for feature importance assessment and influence analysis. The model can support analyzing multimodal feature importance and influence from multi-level views.
    \item Quantitative model performance evaluation, case studies, and expert interviews that demonstrate the effectiveness and usability of \systemname.
\end{itemize}

% based on the real-world dataset collected from 1,172 children in Hong Kong. The 

\section{Related Works}

In this section, we reviewed the related work on visualization for \textcolor{black}{time-series data,} healthcare, and model interpretation and visualization.

    % \subsection{\textcolor{black}{Visualization for Time\-Series Data}}

\subsection{\texorpdfstring{\textcolor{black}{Visualization for Time-Series Data}}{}}

\textcolor{black}{Visualization techniques are important for analyzing time-series data, such as sensor data and meteorological data. Line charts and bar charts are commonly used to visualize single or multiple time series, with time as one dimension and time-dependent attributes as another. For example, \emph{ComVis} developed by Matkovic et al. \cite{matkovic2008comvis} employed curve view and color line view to depict the families of time series data. Recent advancements in interactive visualization techniques have enabled the exploration of time-series datasets with varying scales and contexts. Mansoor et al. \cite{mansoor2021visual} developed an interactive visual analytics tool to discover incorrect or missing annotations by visualizing context-aware human behaviors and unlabeled data chunks. Furthermore, visual analytics techniques have been integrated with statistical and machine learning models to provide deeper insights into time-series patterns and support decision-making. In \cite{xu2021mtseer}, an interactive visual exploration tool was developed for multivariate time-series forecast models.} 

% Interactive visualization techniques have advanced in recent years, enabling the exploration of time-series datasets with varying scales and contexts. Mansoor et al. \cite{mansoor2021visual} developed an interactive visual analytics tool to discover incorrect or missing annotations by visualizing context-aware human behavior and unlabeled data chunks. Integration of visual analytics with statistical models, machine learning algorithms, and time-series forecasting enhances insights and supports data-driven decision-making. In \cite{xu2021mtseer}, an interactive visual exploration tool was created for multivariate time-series forecast models.

\textcolor{black}{However, none of these works support comparing time-series data with other modalities. In this study, we use interactive line graphs to present motion patterns at customized time granularities and ranges and ordered stacked bars to directly compare the importance of motion and context patterns.}

% None of these techniques support comparing time-series data with other modalities. In this study, we use interactive line graphs to display motion patterns at customizable granularities and time frames, and ordered stacked bars to directly compare the importance of motion and context patterns.

% \subsection{\texorpdfstring{\textcolor{black}{Visualization for Cohort Comparision}}{}}

    \subsection{Visualization for Healthcare}
    %discuss the model/method used for health profiling/inference
    
    %healthcare-related visualization (Health-Related Data Visualization: tabular data visualization, time series visualization)
    Visualization tools are widely used in the medical field to analyze and interpret complex health data \cite{bernard2018using, zhang2013five, loorak2015timespan, hakone2016proact}. Bernard et al. \cite{bernard2018using} proposed a technique for segmenting and aggregating patient histories to show longitudinal changes. Zhang et al. \cite{zhang2013five} utilized the Five Ws concept to represent patient data.
    Specific disease-focused tools include TimeSpan \cite{loorak2015timespan} for optimizing stroke treatment time and PROACT \cite{hakone2016proact} for localized prostate cancer.
    Visualization tools have also been utilized for developing medical-relevant software \cite{elshehaly2020qualdash} and managing health communities \cite{kwon2015visohc}. 
    VisOHC \cite{kwon2015visohc} is a visual analytics tool for online health community administrators to understand conversation dynamics by visualizing individual OHC conversation threads as collapsed boxes. Furthermore, the use of machine learning to analyze health data has become increasingly popular.
    In \cite{guo2018visual}, a neural network model was applied to identify and segment semantically meaningful progression stages.
    However, these works mainly focus on individual data visualization, lacking the analytics for a large-scale population.

    % Visualization tools are widely used in the medical field to analyze and interpret complex health data \cite{bernard2018using, zhang2013five, loorak2015timespan, hakone2016proact}. For example, Bernard et al. \cite{bernard2018using} proposed a technique for segmenting and aggregating patient histories to show longitudinal changes. Zhang et al. \cite{zhang2013five} introduced a framework utilizing the Five Ws concept for healthcare information visualization. Specific disease-focused works include TimeSpan \cite{loorak2015timespan}, a tool for optimizing treatment time for stroke victims, and PROACT \cite{hakone2016proact}, a visualization tool for prostate cancer patients. Visualization tools have also been used for software development \cite{elshehaly2020qualdash} and managing health communities \cite{kwon2015visohc}. Machine learning has also been applied for health data analysis \cite{guo2018visual}. However, these works mainly focus on individual data visualization and lack analytics for large-scale populations.

    \textcolor{black}{Besides, many visual analytics tools have been developed for the identification and comparison of patient cohorts. Eckelt et al. \cite{eckelt2022kokiri} presented a visual analytics approach that identifies patient cohorts based on user-selected data, ranks attribute importance, and visualizes cohort overlaps and separability. Scheer et al. \cite{scheer2022visualization} reviewed the visualization techniques for time-oriented data in healthcare, supporting the comparison of single patients or cohorts. However, existing approaches mainly focus on the evaluation of attributes for differentiating cohorts and comparison of cohorts, lacking analysis of the importance and influence of multimodal features on specific cohorts.}
     
    Cheng et al. \cite{cheng2021vbridge} proposed VBridge, a visual analytics tool that helps clinicians better understand machine learning model predictions and provides explanations using SHAP values. VBridge was evaluated on a large dataset collected from 12,000 paediatric patients. However, their analytic tool is only compatible with tabular data. The prevalence of wearable devices (e.g., smart watches) makes it feasible to collect continuous signs of users, providing us with new opportunities for healthcare. In this work, we develop a visualization system that enables interactive exploration of the relationships between children's multimodal features and their health profiles across multiple levels. 

    % However, previous healthcare-related visualization research has mostly focused on visualizing electronic health records \cite{zhang2013five, hakone2016proact}, temporal events for patients \cite{loorak2015timespan,bernard2018using, guo2018visual}, and healthcare dashboards \cite{elshehaly2020qualdash, kwon2015visohc, zhuang2022framework}.
    % Few work has explored children's health profiling using both context and motion patterns from multi-level views.
    % In this work, we develop a visualization system that enables interactive exploration of the relationships between children's multimodal features and their health profiles across multiple levels. 

    \subsection{Model Interpretation and Visualization}

    % model interpretation, importance feature selection
    % visualization \cite{cheng2021vbridge} \cite{wang2021m2lens}
    Machine learning and deep learning models have achieved great performance in health-related studies. The interpretation of models can provide insights into the underlying factors affecting children's health and aid in the development of targeted interventions to improve health outcomes. Model interpretation techniques can be categorized into two groups: intrinsic interpretability and post-hoc interpretability, according to whether the model is tailed with interpretability in training.

    Intrinsic interpretability means that the models are inherently interpretable without additional tools \cite{surveyxai}. Examples include combining sequence models and prototype learning for interpretable representations \cite{kddprototypelearning} and using attention mechanisms to identify influential visits and clinical variables \cite{nipsRETAIN}. The gate mechanism is often used to personalize network parameters for different users \cite{ma2019hierarchical}. Studies by Huang et al. \cite{huang2020gatenet} and Chang et al. \cite{chang2023pepnet} leverage gates to adjust feature weights and provide insights into feature importance.

     Post-hoc interpretability interprets models after the decision-making process, including model-specific and model-agnostic approaches. Model-specific explanations are tailored to specific ML models, spanning from shallow ML models \cite{bayesian, kddensembles} to deep learning models \cite{aclrnn, iclrbayes}. Model-agnostic techniques, such as SHAP \cite{NIPS2017shap}, are designed to be plugged into any model disregarding its internal representations. Additionally, algorithms like LIME \cite{kddlime} build local approximated models based on neighboring instances to explain model predictions.

    Various visualization systems have been developed for model interpretation and interaction. ModelTracker \cite{modeltracker} is a versatile interactive visualization tool for performance analysis and debugging in machine learning. INFUSE \cite{tvcgINFUSE} aids analysts in understanding the ranks of predictive features across different feature selection algorithms. RetainVis \cite{tvcgretainvis} is a visual analytics tool designed for electronic medical records in the context of prediction tasks.

    However, the aforementioned methods are designed for one modality. Wang et al. \cite{tvcgm2lens} proposed a visual tool providing understanding and diagnosis of multimodal models for sentiment analysis. Still, there is a research gap in the studies for a combination of two important modalities related to children's health, \ie, tabular and sensor data. In this work, we propose a multimodal learning model with a gate mechanism that can adjust feature weights intrinsically and provide post-hoc analysis of feature influence with the visualization system.

\section{Background and System Overview}

In this section, we first describe the dataset and the pre-processing procedures. Then, we introduce the design process of \systemname. Finally, we present an overview of the system.
% to demonstrate the whole pipeline.
% Based on the iterative interviews with domain experts, we formulate three levels of nine requirements to guide the system design. 

    \subsection{Data Description and Pre-processing}

    \textcolor{black}{We collaborated with the researchers from the Department of Paediatrics and Adolescent Medicine at the University of Hong Kong and the Hong Kong Children's Hospital and collected data from 1,172 children in Hong Kong}. This study was approved by the Institutional Review Board of the University of Hong Kong / Hospital Authority Hong Kong West Cluster (IRB number: UW 19-516). Informed written consent was obtained from the parents of the participants. More detailed data collection and processing procedures can be found in \cite{healthinference}. There are 733 females and 439 males, with ages ranging from 6 to 18 (mean: 12.48, standard deviation: 2.21). Physical measurements, including height, weight, and body fat percentage, were recorded. Questionnaires were used to collect information on the children's socioeconomic status, electronic device usage patterns, financial satisfaction, academic performance, sleep patterns, exercise habits, and dietary patterns. Learning modes were also included as an important input feature. Every participant was assigned a wristband with accelerometer and instructed to wear it except during bathing or showering. The mean wear duration is 13.85 days (standard deviation: 6.94). 
    % \footnote{The data were collected during the COVID-19 pandemic. Many schools in Hong Kong experienced temporary closures and adopted different learning modes (\ie, full-day face-to-face classes, half-day face-to-face and half-day online classes, and full-day online classes) \cite{healthinference}.}
    %ActiLife Software 6.13.4 \footnote{\url{https://actigraphcorp.com/support/software/actilife/}} was used to process the acceleration data and generate acceleration counts. 
    We then resample the raw acceleration data into minute patterns. \textcolor{black}{The details of the dataset can be found in Appendix A.}

Besides the above-mentioned data on children's personal and family characteristics, we also leverage \textcolor{black}{scales that have been widely used in clinical trials and population studies worldwide to measure children's physical and mental health, based on which we derived the following six health indicators for health profiling.}

\textbf{Physical activity intensity (\emph{MVPA}).} Moderate-to-vigorous intensity physical activity has great benefits on children's health. It is also an important indication for children's health \cite{world2010global}. 

\textbf{Physical functioning (\emph{PHYF}).} Physical functioning measures the ability to perform basic and instrumental activities in daily life \cite{garber2010physical}, measured by the Physical Functioning Scale in PedsQL model \cite{varni2005pedsql}.

\textbf{Health confidence (\emph{VVAS}).} The health confidence is measured by a visual analogue scale from EQ5D \cite{herdman2011development}, indicating the generic quality of life. It is an important index of future morbidity and mortality \cite{jylha2009self}.

\textbf{Psychosocial functioning (\emph{PSYF}).} Psychosocial functioning is the ability to cope with stress, form relationships, and develop identity, measured by the PedsQL model \cite{varni2005pedsql}.

\textbf{Resilience (\emph{RESI}).} Resilience is the ability to adapt and cope with adversity, stress, or trauma, promoting positive outcomes and well-being \cite{masten2002resilience}, measured by the Connor-Davidson Resilience Scale \cite{connor2003development}.

\textbf{Connectedness (\emph{CONN}).} Connectedness measures the children's sense of belonging and connectedness to their families and teachers, measured by the modified Resnick Social Connectedness Scale \cite{rew2001correlates}.

After data pre-processing procedures, we obtain the context and motion features for the 1,172 participants and their corresponding six-dimensional health indicators for health profiling.

    \subsection{Design Process and Requirements}
    \label{sec:design_process}
% Our goal is to support children's health profiling based on context and motion characteristics and also present an in-depth and systematic exploration of the relationships between context and motion characteristics and health profiles. To this end, w

    \textcolor{black}{The goal of the system is to help researchers in associated domains (\eg, public health, medicine, and data science) to explore health profiling with large-scale multimodal data and gain new insights}. We interviewed eight experts (E1-E8) with experience in data analytics related to the health or medical domain and three experts (D1-D3) knowledgeable in visual communication and design. E1 and E2 are postgraduates studying public health. E3 and E4 are postgraduates knowledgeable in obstetrics and gynecology. E5 is working on the medical Internet of Things. E6 is a university research assistant in \textcolor{black}{adolescent} medical-related fields. E7 is a medical laboratory physician. E8 is a postgraduate knowledgeable in data analytics for mental health \textcolor{black}{of children}. D1 and D2 have been working on product design and UI design, respectively. D3 is a postgraduate knowledgeable in the design of Internet of Things platforms. Based on their feedback, we derived a set of requirements to guide the design of the system prototype. In the later stages, we discuss with these experts iteratively to update the system. We summarized the feedback and formed them into the following design requirements.

    \textbf{R1. Enable \textcolor{black}{efficient} presentation of multi-modal input features from multiple levels.} In this work, there are two modalities of input data (tabular and sensor data). E1 and E4-E6 hope the system can present these data \textcolor{black}{efficiently}. D1 said that data with different modalities usually have different suitable graphs; mixing this information together may confuse users and require users to pay extra effort to understand. D2 added that, some traditional graphs that have a long history and are frequently used in our daily life, are easier for users to grasp information quickly. E3 and E8 wished to explore the multimodal input features from different levels (\ie, overall, group, and individual).

    \textbf{\textcolor{black}{R2. Provide an interactive analysis environment to explore the correlation between different input features.}} There is a wide variety of context features. Presenting the correlation between these features can help researchers identify patterns and relationships between context features, which can provide insights into the underlying factors that affect children's health status. It can also help the development of targeted interventions to improve health outcomes. Many interviewed experts (E1-E3 and E6-E8) mentioned that they want to see the correlation of features in an \textcolor{black}{efficient} way. \textcolor{black}{To meet this requirement, the system should provide an interactive analysis interface for the correlation.}

    \textbf{R3. Facilitate the exploration and comparison of health profiles from multiple levels.} Visualizing a comprehensive picture of health profiles can help users explore the data distribution and characteristics quickly. For example, E1 and E2 said the overview visualization would be helpful in describing the population characteristics in their research area. E5 mentioned challenges in identifying common patterns with only an individual view due to significant differences between individuals. Besides, as mentioned by D4 and E7, the system should present a multi-dimensional health profile for each individual \textcolor{black}{effectively} for in-depth investigation. D1-D3 pointed out that radar charts are a common and useful visualization technique for multi-dimensional indicators. E7 also hopes the system can facilitate an \textcolor{black}{efficient} comparison of health profiles between different individuals. 

    \textbf{R4. Support quick screening and comparison between the participants according to their gender and age groups.} Gender and age are critical in health-related analysis, and it is crucial to support population screening with different gender and age groups. Researchers commonly compare behaviors and health outcomes between different genders and age groups in their research, as noted by E1 and E2. According to E3, health disparities can arise due to gender or age, resulting in inequitable health outcomes. Additionally, E7 stated that different gender and age groups often have different health-related baselines.

    \textbf{R5. Facilitate the exploration of feature importance and influence for each health indicator from multiple levels.} Analyzing the importance and influence of input context and motion features facilitates the identification of key predictors, which can help researchers focus their efforts on the most influential factors and develop targeted interventions to improve health outcomes (E1, E2, E6, and E8). It can also help researchers interpret model results and make their analysis more practical \cite{lundberg2017unified}. By eliminating irrelevant or redundant features, the model can be optimized \cite{duangsoithong2009relevant}. Furthermore, since there exist disparities between individuals of different genders, ages, etc., some experts also hope the system can present the personalized feature importance and influence for each individual (E3 and E4).

    % \textbf{\textcolor{black}{R6. Support interactive and comprehensive analysis from multiple levels.}} It is observed that researchers from different associated areas may have different focuses and information needs. E1 and E2 hope the system can focus more on the group level since group patterns are more important in their research domain, while E3 pays more attention to the individual level. Others usually start from the group level but need to investigate individual cases. Experts wished to explore the data with customized settings. The system should be interactive and support comprehensive data exploration for researchers with different focuses. 

    \begin{figure}[!t]
          \centering
          \includegraphics[width=.5\textwidth]{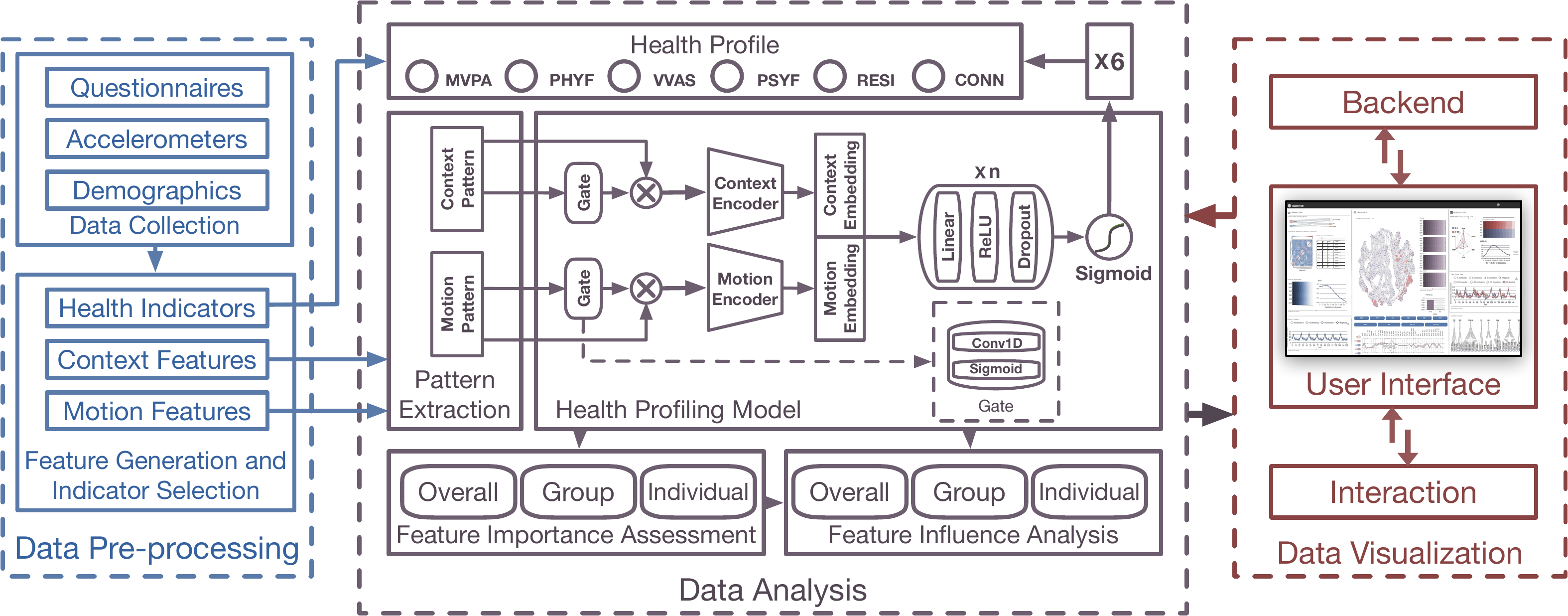}
          \vspace{-0.3cm}
          \caption{\textcolor{black}{Overview of the system and pipeline of \systemname.}}
          \label{fig:system}
            \vspace{-0.5cm}
      \end{figure}

\subsection{System Overview}
\figurename{~\ref{fig:system}} presents an overview of the \systemname system. It consists of three modules, including a data pre-processing module, a data analysis module, and a data visualization module. In the data pre-processing module, the large-scale questionnaire, demographics, and accelerometer personnel raw data are transformed into context features, motion features, and health indicators for each individual stored in the database. In the data analysis module, the motion and context features are extracted into motion and context patterns. A health profiling model based on multimodal learning is used to infer the six-dimensional health indicators \textcolor{black}{(\emph{MVPA}, \emph{PHYF}, \emph{VVAS}, \emph{PSYF}, \emph{RESI}, \emph{CONN})}. Then, based on the model, the importance and influence of features are further analyzed. The backend part is implemented using Python and PHP. In the data visualization module, a front-end web application is built based on Vue.js \cite{vue} and ECharts.js \cite{echarts} with multiple coordinated views to support a comprehensive analysis.

\section{Data Analysis}

% We present an interactive visual analytic system \systemname (\textcolor{black}{\figurename{~}}) to aid researchers in associated domains in exploring children's physical and mental health profiles with multimodal data. In this section, we first present the overview of the system architecture. Then, we illustrate the data analysis pipeline in detail. Finally, we describe the visual design of the system.

In this section, we introduce the data analysis pipeline. First, we extract context and motion patterns. Second, a multimodal learning model is used to profile children's health from six perspectives. Based on the model, we analyze the multimodal feature importance and influence.

\subsection{Pattern Extraction}

We extract context and motion patterns for each participant. The context pattern is extracted from children's context features. \textcolor{black}{In this study, there are 47 different context features, detailed in Appendix A.} $K$-nearest neighbor algorithm is used for data imputation \cite{malarvizhi2012k} and then Min-Max Scale \cite{jayalakshmi2011statistical} is used for data normalization. We further use one-hot encoding to process the categorical features, including gender \textcolor{black}{(binary gender)} and learning modes \textcolor{black}{(three learning modes)}. \textcolor{black}{Thus, we obtain a $1\times 50$ context pattern for each participant}. The motion pattern is extracted from the tri-axial accelerometer data collected using wristbands. More specifically, we aggregate and average the values for each minute in a week to generate the weekly pattern (a sequence with $7\times 24\times 60=10080$ entries). Here we extract the pattern in a weekly manner according to the mean and standard deviation of participants' wear duration. In this way, for tri-axial accelerometer data, we obtain a $3\times 10080$ motion pattern for each participant. The number of channels can be extended when there are other signals.

        % \subsubsection{Tabular Data}
        
        % \subsubsection{Time series Data}

    % \subsection{Health Profiling Indicators}

    \subsection{Health Profiling Model}

     The health profiling model (\figurename{~\ref{fig:system}}) consists of two streams, including two gates, one context encoder, one motion encoder, a stack of fully connected layers (FCN), the rectified linear unit layers (ReLU), and Dropout layers (Dropout). The gate consists of a 1D convolution layer, a ReLU layer, and a Sigmoid layer \cite{harrington1993sigmoid}. The context encoder consists of a stack of FCN, ReLU, and Dropout. The motion encoder follows the similar structure of the IMU encoder in \cite{mollyn2022samosa, moon2022imu2clip}, consisting of a GroupNorm layer, several blocks of a 1D-CNN and Max Pooling, another GroupNorm layer, and a GRU combining the CNN output and generating the motion embedding. \textcolor{black}{The detailed configuration of the model can be found in Appendix D.}

     More specifically, the context pattern \textcolor{black}{($1\times 50$)} is input into the context gate first. Then, the point-wise products \textcolor{black}{($1\times 50$)} of the context pattern and the output of the context gate \textcolor{black}{($1\times 50$)} are input into the context encoder to generate context embedding \textcolor{black}{($1\times 64$)}. Similarly, the motion pattern \textcolor{black}{($3\times 10080$)} is also first input into the motion gate. The point-wise products \textcolor{black}{($3\times 10080$)} of the motion pattern and the output of the motion gate \textcolor{black}{($3\times 10080$)} are input into the motion encoder to generate motion embedding \textcolor{black}{($1\times 64$)}. Then, the context embedding and motion embedding are concatenated together \textcolor{black}{($1\times 128$)} and input into the stack of FCN, ReLU, and Dropout for inferring the health indicator. A Sigmoid layer is used for inferring the probability of a health indicator being predicted as a normal level. For each indicator, the probabilities will be normalized by the min-max method. The six different health indicators are profiled respectively by six models. 
     
     % , \ie, $y_i=(1+e^{-f(x^{(3)}_i)})^{-1}$

    %  Denote the input features as a set $X^{(0)}$, where $x^{(0)}_i\in X^{(0)}$ is the input feature of participant $i$. The input feature consists of context ($c^{(0)}_i$) and motion ($m^{(0)}_i$) patterns, \ie, $x^{(0)}_i=\{c^{(0)}_i,m^{(0)}_i\}$. These two patterns are first input into two gates respectively and get the output $c^{(1)}_i$ and $m^{(1)}_i$. 
     
    %  We further get the point-wise product of $c^{(0)}_i$ and $c^{(1)}_i$, and the point-wise product of $m^{(0)}_i$ and $m^{(1)}_i$, \ie, $c^{(2)}_i=c^{(0)}_i\bigotimes c^{(1)}_i$, and $m^{(2)}_i=m^{(0)}_i\bigotimes m^{(1)}_i$. Then, $c^{(2)}_i$ is input into a context encoder to generate context embedding, and $m^{(2)}_i$ is input into a motion encoder to generate motion embedding, \ie, $c^{(3)}_i=CE(c^{(2)}_i)$ and $m^{(3)}_i=ME(m^{(2)}_i)$.
    
    % After that, the context embedding and motion embedding are concatenated together, denoted as $x^{(3)}_i$. Then, $x^{(3)}_i$ is input into a stack of FCN, ReLU, and Dropout layers, and a Sigmoid layer is used for inferring the probability of a health indicator being predicted as at-risk level, \ie, $y_i=(1+e^{-f(x^{(3)}_i)})^{-1}$.
    
    % The above process is conducted six times respectively, corresponding to the six health indicators.

        \subsection{Feature Importance Assessment}

         \textcolor{black}{The gate mechanism manages the flow of information by selectively adjusting the interim message through the model, thus influencing the impact of specific features or inputs. In this work, we use gates with learnable parameters to personalize the corresponding weights of input features by increasing the weights of more important features and decreasing the weights of less important features. In this way, we can enhance the model's flexibility and adaptability in capturing complex relationships between multimodal features and multiple health outcomes.} By analyzing the intrinsic output of the gate, we evaluate the personalized feature importance. Furthermore, we aggregate the personalized feature importance to evaluate the overall-level and group-level feature importance.

         The gate consists of a 1D convolution layer with input channels equal to output channels and a Sigmoid layer, through which we assess the feature importance. Both the stride and the kernel size are set to 1. For the context pattern, the input and output channels are both one. For the motion pattern, the input and output channels are equal to the channels of the input time series (three in this case). In this way, the 1D convolution assigns the weight and bias corresponding to each input feature, and the Sigmoid layer will normalize the output of 1D convolution into $[0,1]$. 
         % The input feature of participant $i$ is denoted as $x^{(0)}_i$, and the corresponding output of the 1D convolution layer is denoted as $x^{(0)^{'}}_i$.
        %  $x^{(0)}_i=\{c^{(0)}_i,m^{(0)}_i\}$.
        %  Therefore, the output of the 1D convolution layer of participant $i$ is as follows,
        % \begin{equation}
        %     x^{(0)^{'}}_i=\{W^{(0)}_{ci}c^{(0)}_i+b^{(0)}_{ci},W^{(0)}_{mi}m^{(0)}_{i}+b^{(0)}_{mi}\},
        % \end{equation} 
        % where the $W^{(0)}_{ci}$,  $b^{(0)}_{ci}$, $W^{(0)}_{mi}$, and $b^{(0)}_{mi}$ are the learnable weights and bias for context and motion patterns, respectively. 
        % Then, the Sigmoid layer will normalize the output into $[0,1]$.
        % as follows,
        % $(1+e^{-x^{(0)^{'}}_i})^{-1}$
        % \begin{equation}
        %     c^{(1)}_i=\frac{1}{1+e^{-W^{(0)}_{ci}c^{(0)}_i-b^{(0)}_{ci}}},\quad m^{(1)}_i=\frac{1}{1+e^{-W^{(0)}_{mi}m^{(0)}_i-b^{(0)}_{mi}}},
        % \end{equation}
        % where $c^{(1)}_i$ and $m^{(1)}_i$ are the outputs of gates for participant $i$ with the inputs of context and motion patterns, respectively. 
        % After training, the values of $W^{(0)}_{ci}$,  $b^{(0)}_{ci}$, $W^{(0)}_{mi}$, and $b^{(0)}_{mi}$ have been learned. 
        For participant $i$ with input $x_i$, we use the corresponding output of gates $g_i=\{c_i, m_i\}$ as the personalized feature importance, where $c_i$ and $m_i$ represents the importance of context and motion features, respectively. 

       Specifically, the importance of motion pattern ${m_{i}}$ is further assessed based on a customized time window (denoted by $W$) selected by users in the visualization system. In this study, ${m_{i}}$ has a shape of $3\times 10080$, corresponding to the tri-axial week motion pattern of each participant. Then, the three sequences of ${m_{i}}$ are combined into shape $1\times 10080$ using the Root Mean Square (RMS). After that, an algorithm with $O(T)$ time complexity (detailed in Appendix B) is used to find the time slot with the length of $W$ that has the maximal average importance. The above steps are repeated to find the second, third, ..., etc. most important time slots for motion patterns. The most important time slot in the previous iteration will be removed from the sequence before the next iteration. In this way, we can get the importance of the input motion pattern with a customized time window.

        Based on the personalized feature importance, we obtain the overall and group importance of context feature $j$ or time point $j$ in motion pattern by averaging the personal influence of $j$ on corresponding participants. Similarly, the importance of the motion pattern is further determined by the user-selected time window using the same algorithm. 
        % The group-level feature importance follows the same workflow, where $N$ is the number of participants in each group.

    %     Suppose $c^{(1)}_i = [c^{(1)}_{i0}, c^{(1)}_{i1}, ..., c^{(1)}_{iK}]$, $m^{(1)}_i = [m^{(1)}_{i0}, m^{(1)}_{i1}, ..., m^{(1)}_{iT}]$, where $K+1$ is the number of features in context pattern and $T+1$ is the number of time points in motion pattern. For visualization, we further normalize $c^{(1)}_i$ and $m^{(1)}_i$ using Softmax function as follows,

    %     \begin{equation}
    %         \sigma{(c^{(1)}_{ij})}= \frac{e^{c^{(1)}_{ij}}}{\sum_{k=0}^Ke^{c^{(1)}_{ik}}}, j=0,1,...,K,
    %     \end{equation}
    %     \begin{equation}
    %         \sigma{(m^{(1)}_{ij})}= \frac{e^{m^{(1)}_{ij}}}{\sum_{k=0}^Te^{m^{(1)}_{ik}}}, j=0,1,...,T,
    %     \end{equation}

    % Therefore, $\sum_{j=0}^K\sigma({c^{(1)}_{ij}})$ and $\sum_{j=0}^T\sigma({m^{(1)}_{ij}})$ are equal to one. $\sigma{(c^{(1)}_{ij})}$ is used to measure the personalized importance of the input context feature $j$ for participant $i$. 
        
        \subsection{Feature Influence Analysis}
        
        To evaluate the influence of features on the six health indicators in an \textcolor{black}{effective} way, we adopt a perturbation-based method \cite{kaminski2013stochastic} to provide post-hoc analysis of feature influence with the visualization system.

    Suppose we want to analyze the influence of feature $j$ in context pattern on participant $i$. If feature $j$ is a numerical feature, we first control the values of other context and motion inputs, and change the value of feature $j$ from 0 to 1 (after normalization) to visualize the changes of the six outputs corresponding to the six health indicators. If feature $j$ is a categorical feature, we change it to other categories to see the corresponding changes. Similarly, for motion pattern, to evaluate the influence of value at a duration $t$ with customized time window $W$, we control the values of all other features, and change the value of motion features in duration $t$ from 0 to 1 (after normalization, three axes changes simultaneously) to visualize the changes of the six outputs corresponding to the six health indicators. At the group level, the above operation is conducted on all participants in the group to visualize the average changes in the six outputs.

\section{Visual Design}
% In this section, we first present a user scenario to give an overview of the system. Then, we introduce the details of each view in the \systemname system.
\systemname consists of \emph{Summary View}, \emph{Group View}, and \emph{Individual View}, enabling interactive exploration of multimodal features and health profiles. We first \textcolor{black}{present a user scenario to give an overview of the system} and then elaborate on the visual designs of each view.

% In this section, we first introduce a user scenario to illustrate how \systemname helps to explore children's health profile with multimodal features, and then discuss the details in each view. Finally we introduce the  cross-view interactions.

    % \subsection{General Design Scheme}
    % All views follow the same color encoding scheme \textcolor{black}{where ... represents}. 

    % R8, R10

    % \subsection{User Scenario}
    
    % We describe how a researcher in public health uses \systemname to explore the children's health profile with multimodal features. The researcher first inspects the number of participants and the distribution of gender and age (\figurename{~\ref{fig:teaser}}-A1, B1, B3). Then, the researcher specifies the 

    \subsection{\texorpdfstring{\textcolor{black}{User Scenario}}{}}

    \textcolor{black}{We present a user scenario to illustrate how \systemname~facilitates the exploration of multimodal feature importance and influence on children's health profiles. The expert first inspects the \emph{Summary View} (\figurename{~\ref{fig:teaser}-A1}) to gain an overall understanding of the participants, including the distribution of genders and learning modes. She then brushes the correlation heatmap (\figurename{~\ref{fig:teaser}-A2}) and identifies feature pairs with high correlation. By clicking on the grids in the heatmap, the pairs of interest are added to the interactive table for further exploration. Next, the expert explores the Feature Importance (\figurename{~\ref{fig:teaser}-A3}), which displays the top ten most important features for each health indicator. She clicks on the stacked bar to explore the overall influence of the various features on the health indicator. 
    The Motion Feature (\figurename{~\ref{fig:teaser}-A4}) reveals the overall motion patterns. The expert uses the sliders to adjust the granularity and range of time (\figurename{~\ref{fig:teaser}-A4-1, A4-2}). 
    After the overview, the expert goes to the \emph{Group View} (\figurename{~\ref{fig:teaser}-B}) to inspect the important features of different groups (\figurename{~\ref{fig:teaser}-B2}). Using a toggle switch, she compares the context and motion patterns of different groups (\figurename{~\ref{fig:teaser}-B3-1, B3-2}).
    After that, the expert selects the groups and indicators of interest using the buttons (\figurename{~\ref{fig:teaser}-B1-4}). The network graph changes according to her selection (\figurename{~\ref{fig:teaser}-B1-3}), and she switches to a tabular view to browse the detailed health profiles (\figurename{~\ref{fig:teaser}-B1-2}). From the graph and table, the expert finds several individuals of interest. She clicks on the corresponding nodes in the graph or inputs the corresponding IDs in the search box (\figurename{~\ref{fig:teaser}-B1-1}), and then the information in the \emph{Individual View} (\figurename{~\ref{fig:teaser}-C}) changes accordingly. She inspects the health profile of the selected individual (\figurename{~\ref{fig:teaser}-C1}) and explores the personalized feature importance and influence in detail (\figurename{~\ref{fig:teaser}-C2}). Next, she identifies critical features and turns to the Individual Motion Feature and Context Feature part (\figurename{~\ref{fig:teaser}-C3, C4}) to look into the feature patterns. Afterward, the expert gains insights into the influence of specific features and selects the next individual of interest for inspection and comparison.}

    \subsection{Summary View}
    
    The \emph{Summary View} (\figurename{~\ref{fig:teaser}-A}) shows the statistics of input features and their overall importance and influence on the health profiles. It consists of four parts: (1) Categorical Feature Summary (\textbf{R1}); (2) Feature Correlation Analysis (\textbf{R2}); (3) Overall Feature Importance and Influence Analysis (\textbf{R5}); and (4) Motion Feature Summary (\textbf{R1}). 
    
    % \begin{figure}[!t]
    %       \centering
    %       \includegraphics[width=.45\textwidth]{figs/summary_view.png}
    %       \caption{The visual design of \emph{Summary View}. (A) The Sankey diagram depicts the flow between different categorical features. (B) An illustration of using the heatmap and dynamic table to explore feature correlation. In (C), a group of stacked bars  represents the ordered most important ten features of each indicator. (C1) is a slider for customized time window selection. (C2), (C3), and (C4) are illustrations of the influences of categorical context, numerical context, and motion features.}
    %       \label{fig:summary}
    %         \vspace{-0.5cm}
    %   \end{figure}

    \textbf{Description.} In Categorical Feature Summary (\figurename{~\ref{fig:teaser}-A1}), the Sankey diagram depicts the flow between different categorical features (i.e., gender and learning modes) (\textbf{R1}). Tooltips are used to provide values for each category and flow (\figurename{~\ref{fig:teaser}-A1-1}). \textcolor{black}{Two lists are used to customize the categorical features of interests (\figurename{~\ref{fig:teaser}-A1-2}).}
    In Feature Correlation Analysis (\figurename{~\ref{fig:teaser}-A2}), a heatmap presents the Spearman's correlation between numerical context features where red means stronger correlation and blue represents weaker correlation \textcolor{black}{(absolute values)}. In \figurename{~\ref{fig:teaser}-A2}, users can inspect the details (i.e., correlation coefficient $\rho$ \textcolor{black}{with positive or negative signs indicating the positive or negative correlation} and $p$-value) by hovering over the heatmap (\textbf{R2}). The table on the right of the heatmap displays the top $n$ pairs of different features with the strongest correlation, and users can adjust $n$ using the slider over the table. By clicking on a grid in the heatmap, the corresponding pairs will be added to the table. Users can add and delete the pairs in the table freely to compare the pairs \textcolor{black}{of interest}. The table is scrollable, allowing users to view as many pairs as they choose.

    In Overall Feature Importance and Influence Analysis (\figurename{~\ref{fig:teaser}-A3}), stacked bars represent the top 10 most important features for each health indicator (\textbf{R5}). The weights of the ten features are first divided by their sum and multiplied by 100\% for normalization and then sorted by values. As shown in \figurename{~\ref{fig:teaser}-A3}, the most important one among the ten features is on the left (dark blue), and the least one is on the right \textcolor{black}{(pink)}. 
    Since motion feature importance is influenced by the time window, a slider is provided for the customized time window selection (\figurename{~\ref{fig:teaser}-A3-1}). 
    By clicking on the bar, the influence of the corresponding feature on the health indicator will be displayed on the right (\textbf{R5}). Specifically, if the feature is a numerical context or motion feature, the influence will be displayed with a line chart. 
    If it is categorical, the influence will be displayed with a bar chart (\figurename{~\ref{fig:teaser}-A3-2}). The color of the influence chart corresponds to the color of the feature selected. The x-axis represents the feature value \textcolor{black}{with its importance rank}, and the y-axis represents the corresponding health indicator inferred.
    In Motion Feature Summary (\figurename{~\ref{fig:teaser}-A4}), the x-axis is the time based on the time window selected using the slider \textcolor{black}{and the range determined by the slider with two handles below the chart (\figurename{~\ref{fig:teaser}-A4-1})}. The y-axis represents the normalized tri-axial acceleration and their magnitude (\textbf{R1}). By clicking on the legend, users can select the data \textcolor{black}{of interest}.
    % When it is a motion feature, its corresponding time duration in a week will be displayed. 
    
    \textbf{Justification.} Initially, we used pie charts and tables to display categorical information, but experts desired more detailed category flows, so we opted for a Sankey diagram. Due to numerous numerical features, a heatmap was inadequate for comparing interesting pairs, so we implemented a dynamic table to customize pairs of interest. For feature importance, we tried separated bar charts. However, it occupied too much space, and experts wished to see the ranks of importance \textcolor{black}{directly}. Therefore, we sort the feature importance and leverage stacked bars with varying colors to emphasize the more important features. As for the tri-axial acceleration, we use the magnitude while retaining the original three axes for specific inspections.

    %Initially, we used pie charts and tables to display categorical information, but experts desired more detailed and intuitive category flow, so we opted for a Sankey diagram. Due to numerous numerical features, a heatmap was inadequate for comparing and identifying interesting pairs, so we implemented a dynamic table. For feature importance, separated bar charts and different colors were too cluttered, so we sorted the importance ranks and utilized stacked bars of varying darkness. Regarding tri-axial acceleration, we used magnitude while retaining the original three axes for specific inspections.

    \subsection{Group View}

    The \textit{Group View} (\figurename{~\ref{fig:teaser}-B}) shows the statistics of different groups. It consists of three parts: (1) Health Profile Distribution Graph (\textbf{R3}); (2) Group Feature Importance and Influence Analysis (\textbf{R5}); (3) Group Context and Motion Feature Summary (\textbf{R1}).
    
    % \begin{figure}[!t]
    %       \centering
    %       \includegraphics[width=.45\textwidth]{figs/group_view.png}
    %       \caption{The visual design of \emph{Group View}. (A) and (B) are illustrations of networks (A2, B2) with different groups selected using a group of buttons (A3, B3). (A1) and (B1) are legends for nodes and corresponding node numbers. In (C), feature importance and influence of different groups are presented. A toggle switch is used to switch between (D) context features and (E) motion features of each group.}
    %       \label{fig:group}
    %         \vspace{-0.5cm}
    %   \end{figure}
    
    \textbf{Description.} In \textit{Group View}, a group of buttons is provided for screening the health indicators and population groups \textcolor{black}{of interest} (\textbf{R4}). The Health Profile Distribution Graph (\figurename{~\ref{fig:teaser}-B1}) will be updated according to the groups and health indicators selected, as shown in \figurename{~\ref{fig:teaser}-B1-3}. In the graph, each node represents a participant in the group, and the \textcolor{black}{link} represents the similarity of the health profile between two participants (i.e., the Euclidian distance between two profiles). The nodes in the graph are draggable to make sure that each node can be clicked on, even though some nodes have overlapped. For each node, we retain its links to the top ten closest nodes. Note that even though some links of a node are not included in the top ten closest, they might be included in the top ten closest of its neighbors so that each node may have more than ten links. The size and color of nodes are determined by the profile score (\textbf{R3}). The calculation of the score will be introduced in \emph{Individual View}. We obtain the mean ($m$) and standard deviation ($std$) of health profile scores in the selected group and divide the nodes into five divisions \textcolor{black}{according to the 3$\sigma$ rule \cite{huber2018logical}, i.e., the nodes with scores $\geq m$, $m-std\leq$scores$<m$, $m-2std\leq$scores$<m-std$, $m-3std\leq$scores$<m-2std$, and scores$< m-3std$.} \textcolor{black}{The nodes in different groups will have different sizes and colors (the nodes with lower scores have larger sizes and are redder in color}, as shown in \figurename{~\ref{fig:teaser}-B1}). In this way, the individuals with lower scores (worse health status) can be identified quickly. The numbers of participants in the selected groups and each division are displayed at the top left of the \emph{Group View} (\figurename{~\ref{fig:teaser}-B1}). \textcolor{black}{Specifically, besides clicking on the nodes, users can also input the IDs of individuals in a search box (\figurename{~\ref{fig:teaser}-B1-1}) to locate the individuals in the network graph and inspect the corresponding health profiles in \emph{Individual View}. A toggle switch (\figurename{~\ref{fig:teaser}-B1-2}) is used to switch between the network graph and a table presenting the detailed health profile scores of participants with IDs.}
    
    In Group Feature Importance and Influence Analysis, the feature importance and influence of different groups are presented (\textbf{R4}, \figurename{~\ref{fig:teaser}-B2}). It follows the same workflow as that in \emph{Summary View}. In Group Context and Motion Feature Summary, users can switch between context features and motion features by a toggle switch (\textbf{R4}). The context features are presented using a \textcolor{black}{parallel coordinates chart} (\textbf{R1}, \figurename{~\ref{fig:teaser}-B3}). Each y-axis represents one context feature. Users can select the group \textcolor{black}{of interest} (or include all participants) by clicking on the legend (\textbf{R4}) \textcolor{black}{and select the features of interest using the list below the chart (\figurename{~\ref{fig:teaser}-B3-1})}. The motion features of groups are presented using a line graph (\textbf{R1}, \figurename{~\ref{fig:teaser}-B3-2}), which follows the same structure as that in \emph{Summary View}. When the toggle switch is in the motion feature mode, the motion features of each group will be presented using line graphs with a slider for selecting the time window, which follows a similar workflow to that in \emph{Group View}.

    \textbf{Justification.} We initially attempted to use a knowledge graph \textcolor{black}{\cite{cao2020building}} with nodes of participants and indicators, or nodes with rings to represent different health profiles. However, the large number of participants made these designs inefficient to understand. Thus, we used \textcolor{black}{double encoding (i.e., sizes and colors) to distinguish nodes for highlighting critical clusters and individuals}. We tried to use node sizes and colors linearly correlated with health profile scores, but this still resulted in a cluttered design. After several iterations, we divided the nodes with the mean and 3$\sigma$ rule \cite{huber2018logical}, which received satisfactory feedback from experts. 
    % Besides, drawing all links between participants resulted in a cluttered display due to the large number of links. Thus, we retained links to the top ten closest nodes for each node and pruned other less significant links. Each node may have more than ten links, as some links not included in the top ten closest of a node might be included in the top ten closest of its neighbors. 
    To avoid a cluttered display caused by numerous links between participants, we only retained the top ten closest links for each node and pruned less significant ones. A node may have more than ten links if some of its links are included in the top ten closest of its neighbors.
    Additionally, experts found it difficult to compare different groups with heatmaps presenting contextual features correlation for each group. Thus, we changed to present the context features of each group and all participants using a \textcolor{black}{parallel coordinates chart} directly.

    %We attempted to use a knowledge graph with nodes representing participants and health indicators, but it was ineffective and confusing for experts due to the large number of participants. Therefore, we simplified the design using different node sizes and colors, and divided nodes into five groups based on mean and 3$\sigma$ rule. We retained links to the top ten closest nodes for each node, pruned less significant links, and presented context features of each group using a parallel graph. Heatmaps were deemed less useful and difficult to compare by experts.
    
    \subsection{Individual View}
    The individual view (\figurename{~\ref{fig:teaser}-C}) is the visual component to investigate the features and health profiles of individuals. It consists of four parts: (1) Individual Health Profile (\textbf{R3}); (2) Individual Feature Importance and Influence Analysis (\textbf{R5}); (3) Individual Motion Feature Summary (\textbf{R1}); and (4) Individual Context Feature Summary (\textbf{R1}).
    
      %   \begin{figure}[!t]
      %     \centering
      %     \includegraphics[width=.45\textwidth]{figs/individual_view.png}
      %     \caption{The visual design of \emph{Individual View} and its interaction with \emph{Group View}. By clicking the nodes in \emph{Group View} (E), The radar chart (A) will present the health profile of corresponding individuals. Its shape changes with the number of indicators selected (A2). (B) The individual feature importance and influence. (C) and (D) are individual motion and context features.}
      %     \label{fig:individual}
      %       \vspace{-0.5cm}
      % \end{figure}
    
    \textbf{Description.} The users can select a maximum of 2 individuals for direct comparison in the \emph{Individual View}. Two different colors (i.e., dark blue and dark red) correspond to two individuals (\textbf{R3}). In Individual Health Profile (\figurename{~\ref{fig:teaser}-C1}), a radar chart presents the multi-dimensional health profile. The profile score is determined by the area of the polygon formed by the health indicator scores (\textbf{R3}). The shape of the radar chart will be changed according to the health indicators selected (\figurename{~\ref{fig:teaser}-C1-1}). Specifically, if the user only selects one indicator, the score is the value of the selected indicator, and when there are two indicators selected, the score is the mean value. The score will be normalized separately, with different health indicators selected. 
    The individual feature importance and influence are displayed on the right of the radar chart (\figurename{~\ref{fig:teaser}-C2}), following the same workflow as those in \textit{Summary View} and \textit{Group View} (\textbf{R5}). 
    When there are two individuals, the feature importance part can be scrolled to see the full content (\textbf{R3}). Also, users can click on the feature \textcolor{black}{of interest} to see its influence on the corresponding indicator of an individual. 
    Specifically, the influence part can also be scrolled, and users can freely add and delete the influence charts (\textbf{R3}, \textbf{R5}). 
    The Individual Motion Feature Summary (\figurename{~\ref{fig:teaser}-C3}) follows a similar workflow to those in \emph{Summary View} and \emph{Group View} (\textbf{R1}). It also allows the direct comparison between two individuals (\textbf{R3}). 
    As for the Individual Context Feature Summary (\figurename{~\ref{fig:teaser}-C4}), we use separated trees to distinguish between different categories of context features. The value of each context feature is directly presented at the \textcolor{black}{leaf} (\textbf{R1}). \textcolor{black}{If the value is imputed, it would be marked with pink and the symbol `*'}. 
    Users can click on the nodes of feature categories to pack up those \textcolor{black}{uninteresting} features (\figurename{~\ref{fig:teaser}-C4-1}) and focus on those \textcolor{black}{of interest}. When there are two individuals, it can also be scrolled down to see the information of the other individual.

    \textbf{Justification.} Initially, we used a bar chart to display the multi-dimensional health indicators. However, experts thought the bar chart is more suitable for emphasizing the difference among indicators. Therefore, we decided to use the radar chart to \textcolor{black}{directly} present the multiple indicators of equal interest. \textcolor{black}{We use the area of the radar chart to represent the health profile score because it is consistent with the visual representation and more sensitive to capture nuanced multidimensional information. Using area highlights individuals with poorer health status, preferred by experts for efficient analysis, avoiding excessive results from mean values.}
    Furthermore, six-dimensional radar charts confused experts by displaying axes of unselected indicators. Therefore, we removed the \textcolor{black}{uninteresting} dimensions and changed the chart shape. For context features, we originally used tables and \textcolor{black}{parallel coordinates charts}, but experts found them less \textcolor{black}{effective} for category information. We then changed to tree charts, which provide a clearer display of contextual features within categories and are easily collapsible.

%Also, experts wished to directly compare two individuals, and thus we make the \emph{Individual View} compatible with the case of two individuals. 

    %We switched from bar charts to radar charts to present multiple health indicators of equal interest, but six-dimensional radar charts confused experts by displaying axes of unselected indicators. Thus, we removed uninterested dimensions and changed the chart shape. We made the "Individual View" compatible with comparing two individuals. For contextual features, we replaced tables and parallel charts with tree charts, which provide clearer displays and are collapsible.

    \subsection{Cross View Interactions}
    In \systemname, we provide various cross-view interactions for users to explore the multiple coordinated views \cite{munzner2014visualization}. 
    
    \textbf{Ubiquitous Tooltips.} Tooltips are provided in all views for detailed information, including the statistic values and explanations of abbreviations. A summary of abbreviations and their corresponding ID and descriptions is also provided at the top right corner of the system. 

    \textbf{Customized Parameter Setting.} In all views, users can choose the time window and range for presenting motion features and analyzing feature importance and influence. The range of time windows is between 5 minutes and 120 minutes, taking 5 minutes as a step.
    
    \textbf{Indicator, Group and Individual Screening.} In \emph{Group View}, users can choose the health indicators, groups, and individuals \textcolor{black}{of interest}. The details of the corresponding individuals will be displayed in the \emph{Individual View}. The shape of the radar chart in \emph{Individual View} also corresponds to the health indicators selected in \emph{Group View}. 
    
    \textbf{Dynamic Contents.} In \emph{Summary View}, we provide an interactive table for users to freely add and remove records. In \emph{Individual View}, users can also freely add influence graphs by clicking on the importance bar charts and remove graphs by the delete buttons. Users can choose a pair of individuals in \emph{Group View} for comparison in \emph{Individual View}.
    
    % \textbf{Scale, Zoom, and Filter} In \emph{Summary View}, \emph{Group View}, and \emph{Individual View}, charts support zooming and scaling for users to investigate the details. Users can also filter the data presented in the graph by clicking the corresponding legends in each chart.

    \textcolor{black}{\textbf{Justification.} We considered a sequential presentation of views. However, after consulting with experts, we found it was time-consuming and inflexible due to frequent revisits of different views. Also, it may increase users' cognitive load for remembering and integrating information from different views. It also limited the potential for dynamic interactions between views. Users might not be able to immediately observe the impact of their actions in one view on other related views. Although presenting all views together increased the visualizations on one page, the workflow and rationale behind many visualizations in different views are interconnected, allowing users to understand the entire workflow by comprehending one view.}

\section{Evaluation}
In this section, we first evaluate the performance of the proposed health profiling model and then evaluate the effectiveness and usability of \systemname with two case studies and expert interviews.

    \subsection{Model Performance}

    \textbf{Settings.} The 1,172 participants are divided into a training set (80\%) and a test set (20\%). \textcolor{black}{The parameters are tuned by grid search and validated through five-fold cross-validation on the training set to select the optimal settings of parameters.} We use the area under the receiver operating characteristics curves (AUC) \cite{brown2006receiver} and calculate the mean (mAUC) of the six health indicators as evaluation metrics. We compare (3) the proposed health profiling model (HPM) with the following interpretative baseline methods: (1) Support Vector Machine (SVM), and (2) XGBoost (XGB). Since SVM, and XGB can only process data from a single modality, for fair comparisons, we conduct experiments under three different settings: (a) using only context pattern, (b) using only motion pattern, (c) using both context and motion patterns (the motion pattern is directly flattened and concatenated with the context pattern for SVM and XGB). Furthermore, we conduct the experiment (d) without gates for the proposed model HPM. \textcolor{black}{All models are deployed on a server with a CPU of AMD EPYC 7302 16-Core Processor and one NVIDIA GeForce RTX 3090.}

     \begin{table}[!t]
      \tiny
      \renewcommand{\arraystretch}{1}
      % \parbox{\linewidth}{
        \centering
        \caption{The model performance results (AUC) and \textcolor{black}{inference time (in seconds)} under different settings: (a) using only context features, (b) using only motion features, (c) using both context and motion features, and (d) using both context and motion features without gates.}
        \label{tab:results}
            \begin{tabular}{p{0.7cm}p{0.3cm}ccccccccccccc}
          \toprule
          \bfseries Model & \bfseries Setting & \bfseries \emph{MVPA} & \bfseries \emph{PHYF} & \bfseries \emph{VVAS} &  \bfseries \emph{PSYF} & \bfseries \emph{RESI} & \bfseries \emph{CONN} & \bfseries mAUC & \textcolor{black}{\bfseries Time}\\
          \midrule

        % (1) LR & (a) & 0.655 & 0.513 & 0.590 & 0.553 & 0.721 & 0.498 & 0.588 \\
        %         & (b) & 0.829 & 0.490 & 0.520 & 0.495 & 0.672 & 0.495 & 0.584 \\
        %         & (c)  & 0.823 & 0.495 & 0.523 & 0.495 & 0.696 & 0.487 & 0.587 \\
        % \midrule
        (1) SVM  & (a)   & 0.696 & 0.674 & 0.680 & 0.700 & 0.874 & 0.629 & 0.709 & \textcolor{black}{0.167}\\
                & (b)  & 0.942 &  0.578 & 0.562 & 0.589 & 0.687 & 0.523 & 0.647 & \textcolor{black}{56.872}\\
                & (c)  & 0.952 & 0.613 & 0.563 & 0.483 & 0.709 & 0.574 & 0.649 & \textcolor{black}{511.78}\\
            \midrule
        (2) XGB  & (a)   & 0.729 & 0.704 &  0.717 & 0.693 & 0.884 & 0.732 & 0.743 & \textcolor{black}{0.038} \\
         &  (b)  & 0.933 & 0.558 & 0.523 &  0.544 & 0.707 & 0.629 & 0.649 & \textcolor{black}{0.334}\\
         & (c) & 0.932 & 0.658 & 0.580 & 0.615 & 0.821 & 0.711 & 0.720 & \textcolor{black}{0.350}\\
            \midrule
          (3) HPM & (a)  & 0.736 & 0.764 &  0.760 & 0.793 & 0.911 & \underline{0.759} & 0.787 & \textcolor{black}{0.001} \\
              (proposed) & (b)  & \underline{0.957} & 0.688 & 0.651 & 0.706 & 0.674 & 0.666 & 0.724 & \textcolor{black}{0.002}\\
                    & (c) & \underline{0.957} &	\underline{0.781} &	\underline{0.785} &	\underline{0.798} &	\underline{0.918} &	0.735 &	\underline{0.840} & \textcolor{black}{0.002}\\
                     & (d) & 0.837 & 0.773 & 0.752 & 0.759 & 0.912 & 0.750 & 0.797 & \textcolor{black}{0.002}\\
          \bottomrule
        \end{tabular}\\
     \vspace{-0.2cm}
    \end{table}

    \textbf{Results.} The proposed HPM model achieves the best mean AUC under all settings (\tablename{~\ref{tab:results}}). Context features are important to all six health indicators, while motion features are effective in inferring \emph{MVPA}. When using only motion features, all models achieve good performance on \emph{MVPA} because acceleration data are significantly correlated with physical activities. HPM achieves good performance on all indicators when using both context and motion features, showing its effectiveness in dealing with multimodal features. The ablation study shows that using both features for HPM is the most effective, and the gate mechanism helps control the weights of multimodal features. The model with the gate mechanism achieves the best or satisfactory performance on all indicators and the best performance on the mean AUC. \textcolor{black}{Besides, the proposed model uses the least amount of inference time.}

        \begin{figure}[!t]
          \centering
          \includegraphics[width=.45\textwidth]{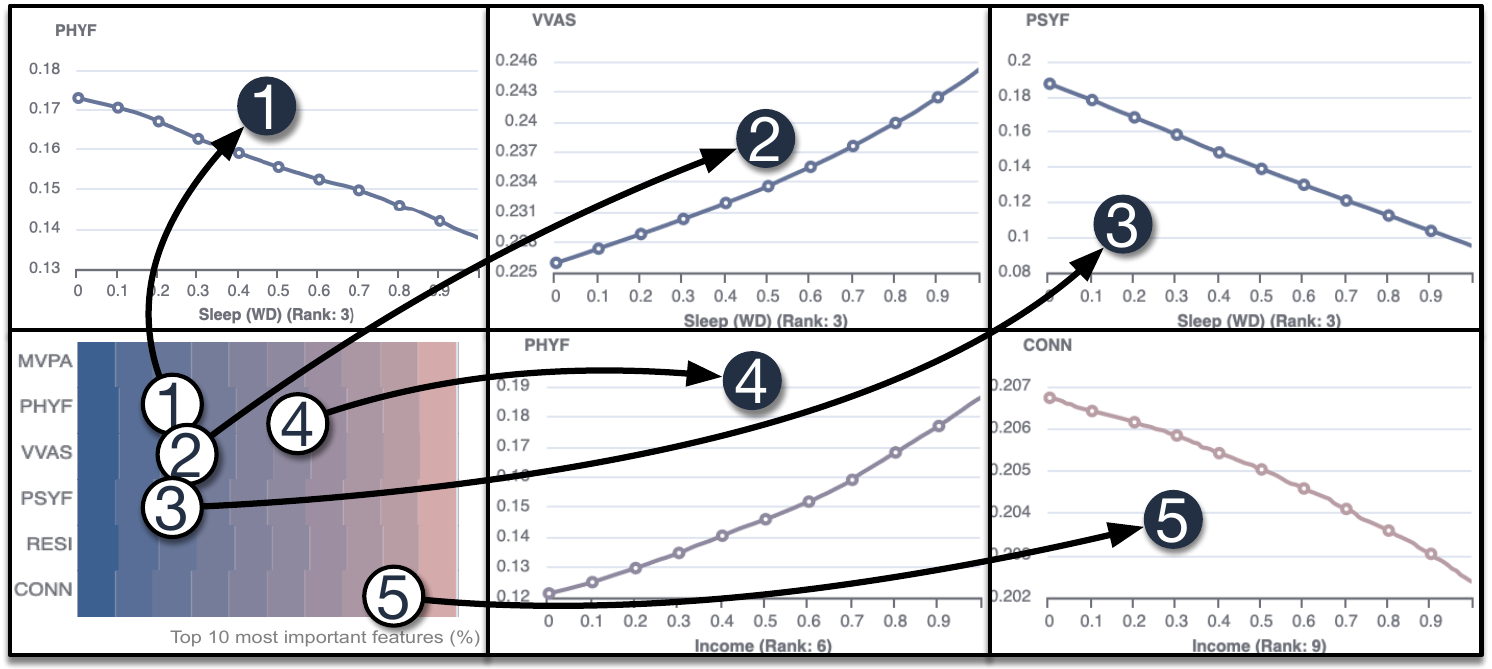}
          \vspace{-0.2cm}
          \caption{An illustration of overall-level inspection. {\large{\textcircled{\small{1}}}-\large{\textcircled{\small{3}}}\large}: Longer sleep duration on weekdays decreases physical functioning and psychosocial functioning but increases health confidence. {\large{\textcircled{\small{4}}}-\large{\textcircled{\small{5}}}}: with the increasing family income, children's \emph{PHYF} increases but \emph{CONN} decreases.}
          \label{fig:case1}
            \vspace{-0.6cm}
      \end{figure}

    \subsection{Case Study}
    \label{sec:case}
    We invited experts (E1-E8) described in Section \ref{sec:design_process} to explore \systemname freely. We summarize their observations and comments to form two cases to demonstrate the features of \systemname.

    \subsubsection{Feature Importance and Influence Investigation}
    With \systemname, experts investigated the importance and influence of various features on the multiple health indicators from the overall, group, and individual levels \textcolor{black}{(\textbf{R5})}. Most experts started by investigating the characteristics of participants. E1 and E2 first explored the \emph{Group View} to explore the number of participants with different genders and age groups (\textcolor{black}{\textbf{R4}}, e.g., \figurename{~\ref{fig:teaser}-B1-3}). They were more interested in the overall and group levels, while E3, E5, E6, and E4 focused more on the individual level. E7 investigated all levels in detail.

        \textbf{Overall-level Inspection.} 
        After exploring the overall feature importance in \emph{Summary View} (\figurename{~\ref{fig:teaser}-A3}), all experts found that motion features had a significant influence on \textcolor{black}{physical activity intensity} and \textcolor{black}{connectedness}, while context features significant influenced \textcolor{black}{physical functioning}, \textcolor{black}{health confidence}, \textcolor{black}{psychosocial functioning}, and \textcolor{black}{resilience} \textcolor{black}{(\textbf{R5})}. E7 thought this phenomenon is consistent with the empirical knowledge since acceleration data directly reflect children's physical activity, and children with higher connectedness with parents and teachers may also have more physical activities with them. E1 found that with the increasing time window, the motion feature would get less important to \textcolor{black}{physical activity intensity} and \textcolor{black}{connectedness}. When the time window was set to 95 minutes, none of the top 10 most important to \textcolor{black}{physical activity intensity} and \textcolor{black}{connectedness} was motion feature \textcolor{black}{(\textbf{R5})}. E1 thus explored the motion feature (\figurename{~\ref{fig:teaser}-A4}) with different time windows (\textcolor{black}{\textbf{R1}}) and thought this might be due to that children usually would not do physical activity for a long time, so when the time window increased, the variation of the importance of different time slots in motion features decrease, leading to the decrease of the importance of motion feature. After exploring the feature influence in detail, E7 was surprised that the sleep pattern showed great importance on \textcolor{black}{physical functioning}, \textcolor{black}{health confidence}, \textcolor{black}{psychosocial functioning}, and \textcolor{black}{resilience} and had different influence on them. \textit{"It is interesting that longer sleep duration on weekdays decreases the physical functioning and psychosocial functioning, but increases health confidence"} (\figurename{~\ref{fig:case1}-\large{\textcircled{\small{1}}}-\large{\textcircled{\small{3}}}}). E2 found that children's \textcolor{black}{physical functioning} increased with family income and thought it was in line with her knowledge, but children's \textcolor{black}{connectedness} decreased when family income increased, which gave her some inspiration (\figurename{~\ref{fig:case1}-\large{\textcircled{\small{4}}},  \large{\textcircled{\small{5}}}}).

    \begin{figure}[!t]
          \centering
          \includegraphics[width=.45\textwidth]{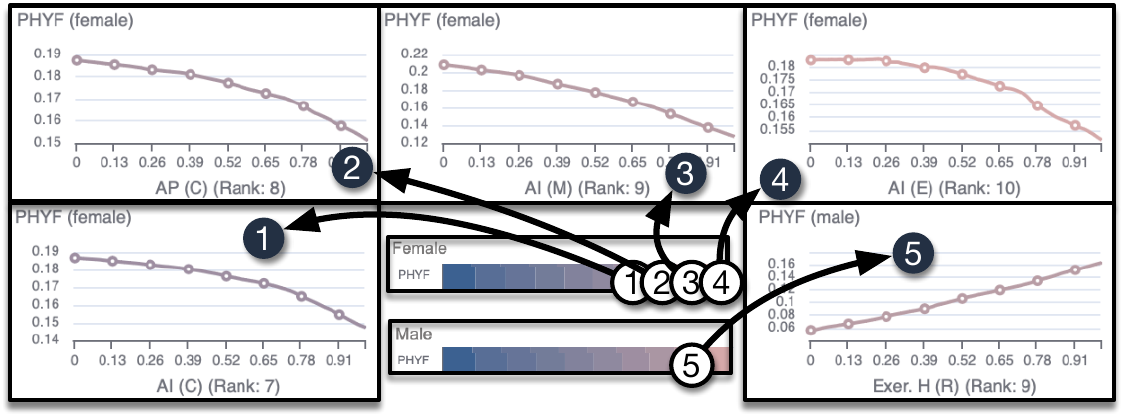}
          \vspace{-0.2cm}
          \caption{An illustration of group-level inspection. {\large{\textcircled{\small{1}}}-\large{\textcircled{\small{4}}}}: The Influence of academic performance and interest on female children's \emph{PHYF}. {\large{\textcircled{\small{5}}}}: The influence of exercise habits of family members on male children's \emph{PHYF}.}
          \label{fig:case2}
            \vspace{-0.3cm}
      \end{figure}

        \textbf{Group-level Inspection.}
        E1 was curious about the differences between female and male groups. After exploring the feature importance and influence in \emph{Summary View} and \emph{Group View} (\textcolor{black}{\textbf{R5}}, \figurename{~\ref{fig:teaser}-A3} and \figurename{~\ref{fig:teaser}-B2}), she found that generally there is no big difference between female and male children, but females have worse \textcolor{black}{physical functioning} than males \textcolor{black}{(\textbf{R4})}. Sleep patterns are important features to \textcolor{black}{physical functioning} for both females and males, while other important features are different. For females, academic performance and interest play an important role, while for males, exercise patterns are important \textcolor{black}{(\textbf{R4})}. Females with higher academic performance and interest tend to have lower physical functioning (\textcolor{black}{\textbf{R1}, \textbf{R3}}, \figurename{~\ref{fig:case2}-\large{\textcircled{\small{1}}}-\large{\textcircled{\small{4}}}}). Males with higher proportions of family members with exercise habits tend to have higher physical functioning (\textcolor{black}{\textbf{R1}, \textbf{R3}}, \figurename{~\ref{fig:case2}-\large{\textcircled{\small{5}}}}). She thought this might be due to the common bias that girls are more gentle and quiet while boys are more active, so parents with exercise habits would schedule more exercise for male children. \textit{"It is important to have good academic performance, but it is a pity that many girls spend too much time on learning and limited time for exercise, which influences their physical functioning. Parents should take it seriously."}
        
        \begin{figure}[!t]
          \centering
          \includegraphics[width=.45\textwidth]{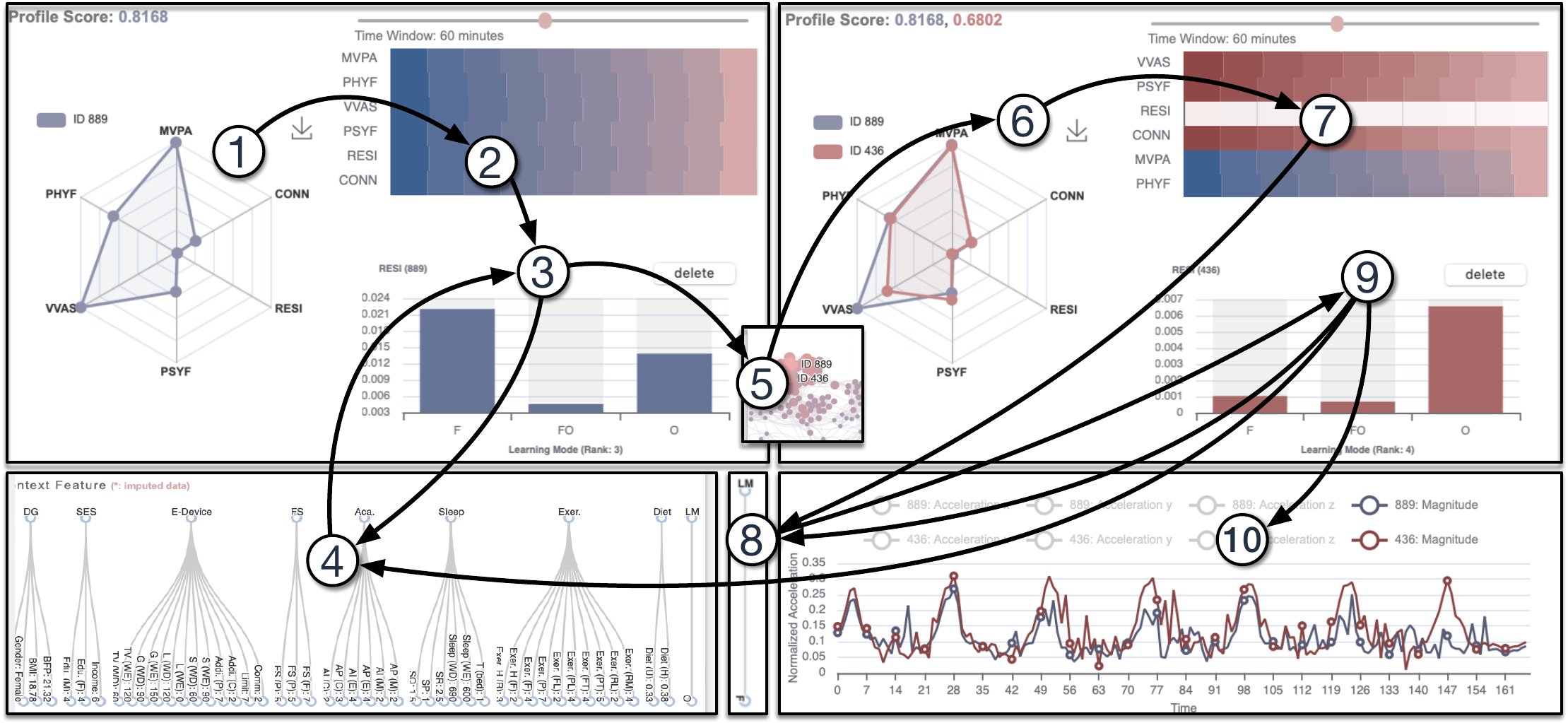}
          \vspace{-0.2cm}
          \caption{An illustration of individual-level inspection. {\large{\textcircled{\small{1}}}-\large{\textcircled{\small{4}}}}: E3 selected an individual (ID 889) and found the learning mode an interesting feature. {\large{\textcircled{\small{5}}}-\large{\textcircled{\small{9}}}}: E3 chose another individual (ID 436) within the same cluster of ID 889 to explore the influence of the learning mode. {\large{\textcircled{\small{4}}}, \large{\textcircled{\small{8}}}}, {\large{\textcircled{\small{10}}}}: E3 further compared the context and motion features of these two individuals.}
          \label{fig:case3}
            \vspace{-0.6cm}
      \end{figure}

        \textbf{Individual-level Inspection.} E3 and E4 browsed some individuals in the significant clusters in \emph{Group View} (\figurename{~\ref{fig:teaser}-B1}) and found that individuals in different clusters have different characteristics of health profiles \textcolor{black}{(\textbf{R3})}. E3 browsed the cluster with low \textcolor{black}{resilience} and selected an individual (ID 889) for further exploration (\figurename{~\ref{fig:case3}-\large{\textcircled{\small{1}}}}). Among the top 10 most important features, she thought the learning mode was quite interesting (\figurename{~\ref{fig:case3}-\large{\textcircled{\small{2}}}, \large{\textcircled{\small{3}}}}). The influence of different learning modes has attracted the attention of many experts, including E3, E5, and E6. They pointed out that they never or rarely thought about the influence of learning modes because children usually always have full-day face-to-face classes. The closure of schools due to the pandemic raises new research questions. E3 further looked at the context feature summary (\textcolor{black}{\textbf{R1}}, \figurename{~\ref{fig:case3}-\large{\textcircled{\small{4}}}}) and found that the child was with full-day online learning mode, while the influence chart showed that the full-day online and online and offline-mixed learning mode would decrease the individual's resilience (\textcolor{black}{\textbf{R1}}, \figurename{~\ref{fig:case3}-\large{\textcircled{\small{3}}}}). She then inspected another individual (ID 436) in this cluster (\figurename{~\ref{fig:case3}-\large{\textcircled{\small{5}}}, \large{\textcircled{\small{6}}}}) and found this individual had a full-day face-to-face learning model (\figurename{~\ref{fig:case3}-\large{\textcircled{\small{7}}}, \large{\textcircled{\small{8}}}}). But different from the individual with ID 889, this individual would have higher \textcolor{black}{resilience} with full-day online and mixed learning modes (\textcolor{black}{\textbf{R3}}, \figurename{~\ref{fig:case3}-\large{\textcircled{\small{9}}}}). \textit{"It seems learning modes have different influences on different individuals."} She thought it was reasonable after comparing their motion and context features (\textcolor{black}{\textbf{R1}}, \figurename{~\ref{fig:case3}-\large{\textcircled{\small{4}}}, \large{\textcircled{\small{8}}}, \large{\textcircled{\small{10}}}}). \textit{"They have very different motion and context features. Since the learning mode would influence many other factors, it is reasonable that the learning mode has different influences on them."}
        
        The case studies demonstrated that \systemname could help experts explore the feature importance and influence from the overall, group, and individual levels effectively and efficiently. They can quickly have an overview of the different influences of multimodal features on different health indicators. It can also give them further insights by comparing different groups and individuals.
        
        \subsubsection{Health Profiles between Different Groups}
\begin{figure}[!t]
          \centering
          \includegraphics[width=.45\textwidth]{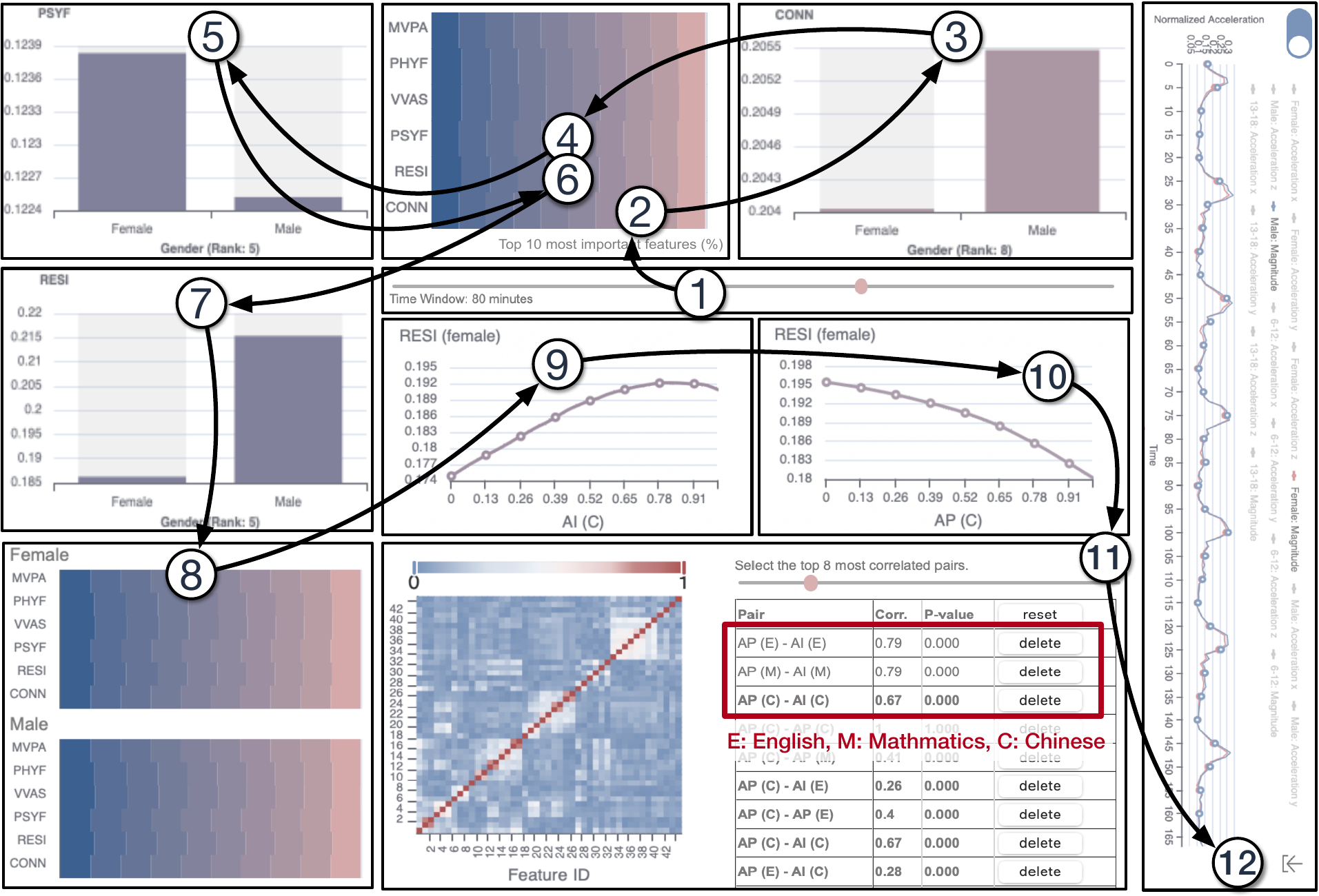}
          \vspace{-0.2cm}
          \caption{An illustration of health profile comparison between female and male children. {\large{\textcircled{\small{1}}}-\large{\textcircled{\small{7}}}}: E8 adjusted the time window and found gender was important to \emph{CONN}, \emph{PSYF}, and \emph{RESI}. {\large{\textcircled{\small{8}}}}: E8 further explore the feature importance of females and males. {\large{\textcircled{\small{9}}}}, {\large{\textcircled{\small{10}}}}: Female's interest and performance on Chinese had different influences on their \emph{RESI}. {\large{\textcircled{\small{11}}}}: E8 used the interactive table to explore the correlation between the interest and performance of three major courses. {\large{\textcircled{\small{12}}}}: E8 used the toggle switch to compare the motion features of females and males.}
          \label{fig:case4}
            \vspace{-0.6cm}
      \end{figure}

        % The comparison of health profiles between different groups of interest 
        Experts can focus on specific health indicators in \systemname. E8 was curious about the difference in mental health between different groups and first went through the overall important features of \textcolor{black}{connectedness}, \textcolor{black}{psychosocial functioning}, and \textcolor{black}{resilience} in \emph{Summary View}. E8 found that motion feature was important to \textcolor{black}{connectedness}, and sleep patterns, dietary patterns, and gender are important to \textcolor{black}{psychosocial functioning} and \textcolor{black}{resilience} \textcolor{black}{(\textbf{R5})}. Then, he clicked on the buttons in \emph{Group View} to remove \textcolor{black}{uninteresting} indicators (\textcolor{black}{\textbf{R3}}) and found there were 61 participants among the 1,172 with the worst level of these three indicators related to mental health. He further clicked on the buttons to control the groups and found that among these 61 participants, 43 were female, and 18 were male \textcolor{black}{(\textbf{R1})}. He was curious about the reason why there were much more female children than male children with worst mental health. Then, he turned back to the feature importance in \emph{Summary View} and found that gender is an important feature to both \textcolor{black}{psychosocial functioning} and \textcolor{black}{resilience} \textcolor{black}{(\textbf{R5})}. As for \textcolor{black}{connectedness}, to look into the influence of context feature, he used the slider to set the time window as 80 minutes and found that gender also became an important feature to \textcolor{black}{connectedness} (\figurename{~\ref{fig:case4}-\large{\textcircled{\small{1}}}, \large{\textcircled{\small{2}}}}). He then clicked on stacked bars to see the influence of genders for the three indicators and found that females tend to have higher \textcolor{black}{psychosocial functioning} but lower \textcolor{black}{resilience} and \textcolor{black}{connectedness} (\textcolor{black}{\textbf{R3}}, \figurename{~\ref{fig:case4}-\large{\textcircled{\small{3}}}-\large{\textcircled{\small{7}}}}). 
        
        Then, he explored the important features of the female and male groups separately (\textcolor{black}{\textbf{R4}}, \figurename{~\ref{fig:case4}-\large{\textcircled{\small{8}}}}), and drew a similar conclusion with E1 that the academic performance of females significantly influenced the female group. But different from E1's conclusion, E8 found females' \textcolor{black}{resilience} increased with their interest in one of the major courses (i.e., Chinese) but decreased with their performance on Chinese (\figurename{~\ref{fig:case4}-\large{\textcircled{\small{9}}}, \large{\textcircled{\small{10}}}}). He thought it was an interesting point and thus turned to the correlation heatmap in \emph{Group View} to see the correlation between children's interest and performance in the three major courses (i.e., Chinese, English, and Mathematics) (\textcolor{black}{\textbf{R2}}, \figurename{~\ref{fig:case4}-\large{\textcircled{\small{11}}}}). He found that compared with English and Mathematics, the correlation between performance and interest in Chinese is lower, which is consistent with our common sense that Chinese is a course difficult to obtain high scores even if you are interested in it and pay great effort (\textcolor{black}{\textbf{R2}}). He thought the reason for the positive influence of interest in Chinese on \textcolor{black}{resilience} might be that those interested in Chinese usually tend to have more extra-curriculum readings. \emph{"Children can learn from how the characters in books overcome challenges and obstacles, which is an excellent way to develop resilience."} As for the \textcolor{black}{connectedness}, since it increased with the increasing motion feature, he used the toggle switch to inspect the motion features of different genders (\textcolor{black}{\textbf{R1}}, \figurename{~\ref{fig:case4}-\large{\textcircled{\small{12}}}}). By comparing the acceleration magnitude, he found that female children have lower acceleration than males almost across the whole week pattern \textcolor{black}{(\textbf{R4})}, which made him draw a similar conclusion with E1 that parents and teachers should encourage female children to do more physical activities.
        
        From this case, we demonstrate the effectiveness of \systemname in exploring the differences in health profiles, feature importance, and influence between different groups.

    \subsection{Expert Interview}
    % do we need to interview experts (\textcolor{black}{xx and xx}) in \ref{sec:design_process}?
    We interviewed \textcolor{black}{two experts (E2 and E3) as described in Section \ref{sec:design_process}} and \textcolor{black}{four new experts (P1-P4)} who used \systemname for the first time to evaluate the usability and effectiveness of \systemname. P1 is a postgraduate in public health. and P2 is a data analyst with a Ph.D. degree. \textcolor{black}{P3 is a Pediatrician, and P4 is a research assistant with an M.Phil. degree studying Paediatrics and Adolescent Medicine.}
    
        \textbf{Settings.} The \textcolor{black}{four} new experts \textcolor{black}{(P1-P4)} were first introduced to the project background and workflows (10 minutes) before exploring the system freely and asking questions for 15 minutes to ensure their familiarity with the system. Then, they started a task-driven exploration with tasks based on case studies in Section \ref{sec:case} for 30 minutes, followed by another 15 minutes of free exploration \textcolor{black}{themselves}. A 30-minute semi-structured interview was conducted with all experts (E2, E3, P1\textcolor{black}{-P4}). \textcolor{black}{The details of the interview including the tasks can be found in Appendix C}. The feedback is summarized below.
        
        \textbf{System Workflow.}
        All experts confirmed the effectiveness of the system workflow of \systemname in profiling children's physical and mental health and analyzing the feature importance and influence. They mentioned that some important features revealed by the system are consistent with their knowledge, and \systemname also provides them with some new insights. \textit{"I have many interesting new findings using the system, such as the different influences of different learning modes"} (E3). Also, \systemname is very helpful in quickly locating and analyzing the groups and individuals of interest. \textit{"I used to take many operations on tables to query and filter the groups \textcolor{black}{of interest}. Using \systemname is much more efficient"} (P1). \textcolor{black}{P3 and P4 appreciated \systemname, especially for the efficiency in identifying critical groups of individuals for further inspection. \textit{``I can quickly figure out the individuals with abnormal health status in the graph and look into the details''} (P3)}.
        E3 pointed out that it was very convenient to use this tool for exploration since she can \textit{"directly look at the context and motion features of critical individuals for in-depth analysis"}. They appreciated the system for considering motion features. \textit{"I used to explore a lot of tabular data, but it is the first time to explore the motion data from wearable devices. It is very inspiring"} (P2). They also like the techniques within the system (e.g., the multimodal deep learning model). E2 mentioned that although she had heard of deep learning many times and was very interested in it, she never tried this technique for exploration because it was difficult for researchers with little knowledge of Computer Science to adopt this technique. \textit{"It was impressive to see the effectiveness of deep learning in handling various data, and it encouraged me to learn related techniques"} (E2). \textcolor{black}{P4 was very impressed by the cross-modality comparison. \textit{``I analyzed both independent and time-dependent attributes but had never compared them together''} (P4).}

        %All experts confirmed \systemname's effectiveness in profiling children's physical and mental health, analyzing feature importance and influence, and quickly locating and analyzing groups and individuals of interest. They found new insights and appreciated the convenience of exploring critical individuals' context and motion features. E3 found the system helpful for exploration, especially in analyzing the motion data from wearable devices. They also liked the multimodal deep learning techniques used in the system, which encouraged E2 to learn related techniques.

        \textbf{Visualization and Interaction.}
        E1 and E2 considered \systemname a comprehensive visual analytics system that fulfilled all design requirements. The \textcolor{black}{four new experts (P1-P4) could easily understand the visual designs after the free exploration and Q\&A and finished the exploration tasks within the allotted time.} P1 and P2 were also impressed by the visualization and interaction. \textit{"The color scheme is beautiful and harmonious"} (P1). At first, they were confused about the abbreviations in the system, but with the help of tooltips, they were able to grasp the meanings of these abbreviations quickly. They agreed that the charts used in the system are easy to understand and compare between different groups and individuals. P1 \textcolor{black}{and P3} were confused by the Sankey chart since they had never used this kind of chart before, but after explanation, they were able to understand it and thought it was a good way to present categorical flows. For feature importance and influence visualization, they thought it was \textcolor{black}{efficient} to see the top 10 most important features in order and explore the corresponding influence by directly clicking on the stacked bars. 
        But P2 thought it would be better to use different colors to distinguish different indicators. E2, E3 \textcolor{black}{and P4} particularly liked the design of dynamic tables for correlation and individual feature influences. They thought it was very useful for comparison. \textcolor{black}{All new experts (P1-P4)} were very impressed by the network graph in \emph{Group View}. \textit{"It directly showed the relationship and clusters of participants. The critical individuals were also significant"} (P1). The usage of radar charts for health profiles was also liked by all experts since it was \textcolor{black}{efficient} and corresponded to the nodes in \emph{Group View}. \textcolor{black}{P3 liked the sliders for customized time granularity and range. \textit{``The system provided substantial interactions so I could explore the data more freely''} (P3).} Experts were impressed by the way views interact with each other, especially for the \emph{Group} and \emph{Individual Views}.

        \textbf{Suggestions.}
        Experts also provide some suggestions based on their exploration. P2 preferred different colors for different time windows. \textit{"I changed the time window, but the bars looked similar to the previous one, making me feel like I failed to change the time window, and I had to confirm that I changed it successfully by looking at the data"} (P2). P1 suggested that although a dynamic table served to display feature correlations, too many features render the heatmap on the left difficult to use. \textit{"The grid in the heatmap is too small, making it difficult to locate the pair I want to explore with mouse"} (P1). \textcolor{black}{P3 hoped the \emph{Individual Views} to support the comparison of more than two individuals. P4 proposed that nodes in the network graph could be removed freely to make the nodes of interest more obvious}.
        E1 suggested that although the tree chart is able to pack up features of no interest, it might be better to directly remove them from the whole system to avoid confusion because their studies usually focus on features from one specific category. E2 preferred to see more important features and suggested allowing users to set the number of important features.

\section{Discussion}
In this section, we discuss the implications, limitations, and generalizability of our work.

    \textbf{Implications.}
    % Many countries around the world are facing a shortage of healthcare resources, especially during the post-epidemic era. 
    Studying the importance and influence of children's personal and family characteristics on their physical and mental health can help provide indications for improving health and help screen the critical population that requires special attention. Besides many traditional attributes, wearable devices (e.g., smart watches) integrated with multiple sensors, such as accelerometers, gyroscopes, and heart rate sensors, can continuously monitor various vital signs of users, and thereby these time series are a significant modality of data for healthcare. The proposed data visualization system is compatible with two important modalities (\ie, tabular and sensor data). We believe \systemname can help researchers in associated domains (\eg, public health, medicine, and data science) to explore health profiling with large-scale multimodal data and gain new insights.
    
    % However, traditional analytical methods may be unable to analyze multimodal data, especially for many researchers in public health and medical areas with little knowledge in dealing with such complex data. 
    % While deep learning techniques have shown their power in handling multimodal data, collaborations between multiple disciplines are encouraged to explore complex data and gain new insights. The proposed analytics system can help researchers in associated domains to freely explore and understand the multimodal data and model.

    %Studying children's personal and family characteristics can improve healthcare and identify those who require special attention. Wearable devices with multiple sensors can continuously monitor vital signs, providing significant healthcare data. However, traditional analytical methods may be unable to analyze such complex data. Deep learning techniques are powerful, but interdisciplinary collaborations are needed. The proposed analytics system enables researchers to explore and understand multimodal data and models in associated domains.
    
    % design implications
    % visualization of a large-scale of the population (critical data)
    % customized configurations
    % We found that the decision-making strategies vary across researchers. Thus, it is important to allow users to adjust the organizational methods of visual information as needed. 

    \textbf{Limitations and Future Work.}
    Although the evaluation demonstrated the effectiveness and usability of \systemname, it still has limitations. First, users may only be interested in parts of features, and \textcolor{black}{the age groups may not be fine-grained enough. Possible improvements include providing a feature checklist and a slider with two handles for age range selection.} 
    % although \systemname supports the filtering of health indicators, population groups, and context features with categories, there are still many input context and motion features with different time windows, while researchers from different domains have different focuses and may only want to pay attention to parts of features. Also, the current age groups may not be fine-grained enough. 
    The views can be updated only to show the analysis results of selected features. Also, the system can be further customized by allowing users to select the number of important features presented. 
    Second, in \emph{Group View}, we prune the less significant links to avoid information overload, which may filter out some useful links. 
    We plan to make the number of links available in the system customized while keeping it within a reasonable range. 
    Third, the system is able to deal with tabular and sensor data. Still, there are many other modalities, such as images and audio. In the future, we plan to involve data with more modalities. 
    Fourth, in the evaluation, although we have confirmed the interpretability of the proposed method with case studies and expert interviews, it might be helpful to compare its interpretability with baselines quantitatively. \textcolor{black}{Fifth, although the current system supports comparison between different gender and age groups, it can be further improved to support the direct identification and comparison of cohorts. Additionally, experts are currently required to undergo training in order to effectively utilize the system. In the future, we plan to allow users to select specific modules to simplify the system, allowing for a more customized experience.}
    Finally, the evaluation only involves experts, while the dataset may also be of interest to a broader range of users, such as parents who want to understand factors influencing their children's health. Additional user studies involving more experts and non-experts are needed to improve the system.

    %Although \systemname demonstrated effectiveness and usability, there are still limitations. First, users may only want to focus on parts of the features, and the age groups may not be fine-grained enough. Possible improvements include providing a feature checklist and a slider for age range selection. Second, some useful links may be filtered out due to information overload. We plan to customize the number of links available. Third, the system can only deal with tabular and sensor data. We plan to involve data with more modalities. Fourth, we need to compare the interpretability of the proposed method with baselines quantitatively. Finally, the evaluation only involves experts, and additional user studies involving more experts and non-experts are needed to improve the system.

        % \subsubsection{The presentation of contextual features} too many features, allow users to select targeted features
        % \subsubsection{The Graph} too many nodes and links 
        % \subsubsection{Multimodal Model Performance} Although the current multimodal model can effectively fuse multimodal features and adjust corresponding weights, it 

    \textbf{Generalizability.}
    We illustrate the generalizability of \systemname from the perspectives of extensibility and transferability.

    First, the system can be easily extended to incorporate more health indicators and diverse demographic groups. For example, there are many other features correlated with children's health status, such as living environment and hereditary factors \cite{karn2003living, kato2005genetic}. By representing these features with tabular or sequence data, they can be easily incorporated into the system. Although we only consider binary gender and two age groups currently following the most existing literature in healthcare domains, it can be easily extended to more diverse genders and fine-grained age groups. Also, other health indicators, such as ADHD and depression, can be incorporated into the health profiles.

    Second, the visual design and analysis pipeline of \systemname can be applied to a wide range of scenarios. For example, it can be applied to the diagnosis of traffic accident hotspots. A road intersection with tabular features (e.g., road width and speed limit) and time series features (e.g., traffic flow and weather conditions) can be classified into a traffic accident hotspot or not \cite{jiang2021crowdpatrol}. The importance and influence of various features on its probability of being a hotspot can be figured out. Similarly, it can also be used to diagnose crime hotspots \cite{zhou2021spatio}, crowded nodes in mobile networks \cite{ma2014opportunities}, and factors increasing the error risk for software and hardware \cite{charette2005software}. The practitioners can first choose dimensions to focus on and then determine the features to be studied. Following the data analysis pipeline, the importance and influence of multimodal features on their tasks can be measured and visualized.
    % The ability of dealing with large-scale, heterogeneous, multimodal data and multi-dimensional indicators makes the system adaptive to many application scenarios. 

    %We show how \systemname is generalizable in terms of extensibility and transferability. The data analysis pipeline and visualization system can be easily extended to incorporate more health indicators and diverse demographic groups. Additionally, the system's visual design and analysis pipeline can be applied to a wide range of scenarios, including traffic accident and crime hotspots, mobile network analysis, and error risk factors for software and hardware. The system's ability to handle large-scale, multimodal data and multi-dimensional indicators makes it adaptable to many application scenarios. Practitioners can focus on specific dimensions and choose relevant tabular and time series features for analysis, measure their impact, and visualize results.

% other tasks related to tabular data and time series data can be studied. 
    
\section{Conclusion}
We introduce \systemname, an interactive visual analytics system for multidisciplinary researchers to explore children's physical and mental health with multimodal features.
% In this paper, we present \systemname, an interactive visual analytics system for researchers from various domains (e.g., public health, medicine, and data science) to explore the correlations between multimodal features and children's physical and mental health profiles. 
Our system utilizes a multimodal learning model with a gate mechanism to profile children's health and support cross-modality feature importance comparison.
% Within the \systemname, a multimodal learning model with a gate mechanism is proposed for health profiling and cross-modality feature importance comparison. 
\systemname has three \textcolor{black}{coordinated} views facilitating the screening of groups and individuals to explore multimodal features and their importance and influence from multiple levels. We demonstrate the effectiveness and usability of \systemname by quantitative evaluation of the model performance, case studies, and expert interviews in associated domains. 

%We introduce \systemname, an interactive visual analytics system for multidisciplinary researchers to explore correlations between multimodal features and children's physical and mental health. Our system utilizes a multimodal learning model with a gate mechanism to provide health profiling and cross-modality feature importance comparison. It facilitates exploring context and motion patterns of individuals, presenting personalized importance and influence of features, and screening groups with different genders and ages. Our system's effectiveness and usability are validated by quantitative evaluation, case studies, and expert interviews in related domains.

% In the future, we are going to improve \systemname from the following directions. \textcolor{black}{First,...}

%% if specified like this the section will be ommitted in review mode
\acknowledgments{%
 % \textbf{For IEEE VIS}, this section may be included in the \textbf{2-page allotment for References, Figure Credits, and Acknowledgments}.
	
	The authors would like to thank Ka-Man Yip, Dr. Hung-Kwan So, Dr. Wilfred H.S. Wong, and Dr. Patrick Ip from the Department of Paediatrics and Adolescent Medicine, LKS Faculty of Medicine, the University of Hong Kong for providing the dataset. We also thank all domain experts for participating this project, and the anonymous reviewers for their valuable comments. This research was supported by the GRF Grant No. 17203320 and 17209822 from Hong Kong, and the start-up grant from The University of Hong Kong.%
}

\bibliographystyle{abbrv-doi-hyperref}
%\bibliographystyle{abbrv-doi-hyperref-narrow}
%\bibliographystyle{abbrv-doi}
%\bibliographystyle{abbrv-doi-narrow}

% \bibliography{template}

%% ^^^^^   FOR IEEE VIS, EVERYTHING HERE MAY BE INCLUDED IN THE    ^^^^^ %%
%% 2-PAGE ALLOTMENT FOR REFERENCES, FIGURE CREDITS, AND ACKNOWLEDGEMENTS %%

% \appendix % You can use the `hideappendix` class option to skip everything after \appendix

% \section{About Appendices}
% Refer to \cref{sec:appendices_inst} for instructions regarding appendices.

% \section{Troubleshooting}
% \label{appendix:troubleshooting}

% \subsection{ifpdf error}

% If you receive compilation errors along the lines of \texttt{Package ifpdf Error: Name clash, \textbackslash ifpdf is already defined} then please add a new line \verb|\let\ifpdf\relax| right after the \verb|\documentclass[journal]{vgtc}| call.
% Note that your error is due to packages you use that define \verb|\ifpdf| which is obsolete (the result is that \verb|\ifpdf| is defined twice); these packages should be changed to use \verb|ifpdf| package instead.

% \subsection{\texttt{pdfendlink} error}

% Occasionally (for some \LaTeX\ distributions) this hyper-linked bib\TeX\ style may lead to \textbf{compilation errors} (\texttt{pdfendlink ended up in different nesting level ...}) if a reference entry is broken across two pages (due to a bug in \verb|hyperref|).
% In this case, make sure you have the latest version of the \verb|hyperref| package (i.e.\ update your \LaTeX\ installation/packages) or, alternatively, revert back to \verb|\bibliographystyle{abbrv-doi}| (at the expense of removing hyperlinks from the bibliography) and try \verb|\bibliographystyle{abbrv-doi-hyperref}| again after some more editing.

\end{spacing}

\end{document}